\documentclass[final,5p,times,pdflatex]{elsarticle}
\usepackage{amssymb,bbold}
\usepackage{amsmath}
\usepackage[utf8x]{inputenc}
\usepackage{natbib}
\usepackage{hyperref}
\usepackage{epstopdf}
\hypersetup{
	colorlinks=true,linkcolor=blue,citecolor=blue,
	filecolor=blue,urlcolor=blue,breaklinks=true
}
\biboptions{sort&compress}
\journal{Physica E}
\RequirePackage{color}

\begin{document}

	\begin{frontmatter}
		\title{Modification of Landau levels in a two-dimensional ring due to rotation effects and edge states}
		\author[ufma]{Lu\'{i}s Fernando C. Pereira}
		\ead{luisfernandofisica@hotmail.com}
		\author[ufma]{Edilberto O. Silva}
		\ead{edilberto.silva@ufma.br}		
		\address[ufma]{Departamento de F\'{i}sica, Universidade Federal do Maranh\~{a}o, 65085-580 S\~{a}o Lu\'{i}s, Maranh\~{a}o, Brazil}
				
		\begin{abstract}
			We investigate the properties of a two-dimensional quantum ring under rotating and external magnetic field effects. We initially analyse the Landau levels and inertial effects on them. Among the results obtained, we emphasize that the rotation lifted the
			degeneracy of Landau levels. When electrons are confined in a two-dimensional ring, which is modeled by a hard wall potential, the eigenstates are described by Landau states as long as the eigenstates are not too close to the edges of the ring. On the other hand, near the edges of the ring, the energies increase monotonically. These states are known as edge states. Edge states have an important effect on the physical properties of the ring. Thus, we analyze the Fermi energy and magnetization.
			In the specific case of magnetization, we consider two approaches. In the first, we obtain an analytical result for magnetization but without considering rotation. Numerical results showed the de Haas-Van Alphen (dHvA) oscillations. In the second, we consider rotating effects. In addition to the dHvA oscillations, we also verify the Aharonov-Bohm-type (AB) oscillations, which are associated with the presence of edge states. We discuss the effects of rotation on the results and find that rotation is responsible for inducing Aharonov-Bohm-type oscillations.
		\end{abstract}
		
		
	\end{frontmatter}
	
\section{Introduction}

\label{intro}

A two-dimensional electrons gas (2DEG) confined in a mesoscopic rings and immersed in an external magnetic field exhibit a variety of physical phenomena \cite{fomin2013physics,Chakraborty2003}. This is due, in part, to the ring geometry itself \cite{fomin2013physics}. On the other hand,  the ring dimensions on the mesoscopic scale have a high influence on the occurrence of physical phenomena. In fact, when the device size is comparable to the mean free path for electrons, or to the phase-coherence length of the electrons, for example, the system enters the mesoscopic scale. In this scale, various physically interesting phenomena occur \cite{BOOK.Heinzel.2008,Book.Murayama.2008,BOOK.Ihn.2010}.
Temperature also has a great influence on the physics of mesocopic systems. Indeed, the phase coherence length of the electron increases significantly at low temperatures \cite{PRB.1988.37.6050,IBM.1988.32.359}.
In quantum rings, we can access
quantum transport phenomena such as magnetoresistance and Hall effect \cite{Inbook.Martins}. The oscillations in magnetoresistance are the solid state analog of the AB effect for electron in vacuum \cite{PRL.1985.55.1610,PRB.1991.44.1657,PRB.1993.48.15148,PRB.2007.76.03510,EPN.2005.36.78}. Quantum rings also exhibit
optical properties. For instance, in Ref. \cite{PRB.2011.84.235103}, the authors study theoretically the optical properties of an exciton in a two-dimensional ring threaded by a magnetic flux. In Ref. \cite{SM.2013.58.94}, the photoionization cross section associated with intersubband
transitions is investigated
threaded by a magnetic flux. Another area of much theoretical and experimental interest is the study of thermodynamic properties in quantum rings. For instance, a quantum ring can carry a persistent current, which is
periodic in the AB flux $\Phi$ with a period $\Phi_{0}= h/e$, the flux quantum \cite{PLA.1983.96.365,FBS.2022.63.58}.
Persistent currents have been detected on mettalic rings \cite{PRL.1990.64.2074,PRL.1991.67.3578} and in experiments on a semiconductor
single loop in the GaAs/GaA1As system \cite{PRL.1993.70.2020}.
Recently, Vasilchenko used the
density functional theory to study the effect of electron-electron
interaction on the persistent currents in
two-dimensional quantum rings containing several electron \cite{CCM.2022.32.e00698}.
Here, we are interested in another thermodynamic property, namely, the magnetization, which in single, isolated
loops it is proportional to the persistent current \cite{PRL.1990.64.2074,PRL.1991.67.3578,PRL.1993.70.2020,PRB.1994.50.8460,PRB.1999.60.5626}. At zero temperature, the magnetization $\mathcal{M}$ is computed as $\mathcal{M}=-dU/dB$, where $U$ is the internal energy.
The magnetization presents an oscillatory behavior as a magnetic field perpendicular to the plane of
the 2DEG changes. These oscillations are known as dHvA effect \cite{Book.Kittel.1987}. The effect is related to the occupation of electronic states at Landau levels. When the magnetic field increases the degeneracy of a Landau level increases, which means that more electrons can be accommodated in the Landau levels \cite{Book.Yoshioka.2002,BOOK.JAIN.2007}. Therefore, in a sample with a fixed number of electrons, increasing the magnetic field strength leads to a reduction in the number of occupied Landau levels. The process continues until all electrons are accommodated in the lowest Landau level. The depopulation of a level and the redistribution of the electrons in the lowest levels lead to the dHvA effect.
The dHvA effect provides one of the best tools for the investigation of the fermi surfaces in metals \cite{Book.Kittel.1987}. The first experimental observation of this
effect was made by de Haas and van Alphen \cite{dHvAeffect.1930}. They measured the magnetization of sample bismuth (Bi) as a function of the magnetic field at $T=14,2$ K and found that the magnetic susceptibility is a periodic function of the magnetic field.

In the 2DEG bulk case Landau, and at zero temperature, the magnetization has sharp, saw-tooth oscillations with a constant
amplitude \cite{ZFP.1933.81.186,SM.1992.11.73,PRB.2002.65.245315,EPJB.2003.36.183}. Nevertheless, a more complete description should take into account the effect of edges on the electronic states of the sample. The edges of the sample can be modeled by hard wall. We then consider a potential that is null inside the sample but that increases abruptly at the edges.
This is the case of the infinite square well potential \cite{SST.1994.9.1305,Book.Yoshioka.2013}. The exact energy of the model cannot be expressed analytically, and has to be determined numerically \cite{PRB.1982.25.2185,PRB.1984.29.1616,JAP.1990.68.3435,PRB.1993.47.9501}. With the above information in mind, we define a quantum ring with inner and outer radii, given by $r_{a}$ and $r_{b}$, respectively. If
\begin{equation}
r_{b}-r_{a} \gg \sqrt{\frac{\hbar}{eB}},     \end{equation}
where the term on the right is the usual magnetic length, then the ring presents the conditions for edge states as well as Landau states to occur \cite{PRB.1993.47.9501}. By using this description for a case of a  quantum dot with $r_{a}=0$ and $r_{b}=R$, Sivan and Imry \cite{PRL.1988.61.1001} shown that the discontinuous jumps in the oscillations are rounded because to edge states are included. Besides, AB oscillations are verified superimposed on the dHvA oscillations. The AB oscillations arise when the Fermi energy is within a bulk Landau level. So, when one edge state crosses the Fermi energy occurs a change in the magnetization, and this leads to the AB oscillations.  In the case of a ring, we point out that these oscillations also occur when the Fermi energy moves into the gap of the Landau levels. In fact, the ring has two edges and the crossing of the states of the outer edge with those of the inner edge at Fermi energy leads to AB oscillations.

Other approaches have been used to investigate oscillations in magnetization as a function of the magnetic field.
For example, Tan and Inkson used an exactly soluble model to study magnetization and persistent current in two-dimensional mesoscopic rings and dots over a wide range of magnetic field strength \cite{PRB.1999.60.5626}. In this approach, the ring is described by a radial potential. Recently, we have used this model to study the effects of curvature on the magnetization and persistent current of rings and quantum dots \cite{AdP.2019.531.1900254,PE.2021.132.114760}. Geometrical effects are relevant in the investigation of systems in the domain of the Solid State Physics \cite{PO.2020.5.100045}.
On the other hand, Maniero et. al. have studied two electrons in a quantum dot under two different confinement conditions. They have reported an oscillatory motion as a function of the magnetic field, which resembles the dHvA
effect \cite{JPB.2020.53.185001,JPB.2021.54.11LT01}. The magnetization and persistent current of two electrons confined by a circular parabolic GaAs quantum dot has been also studied by Nammas using the canonical ensemble approach \cite{JLTP.2020.200.76}.

It is interesting to note that rotating effects have been discussed in the literature \cite{PA.1996.233.503,EL.2001.54.502,PBCM.2006.385.1381,PRB.2011.84.104410,PRA.2014.89.063620,AofP.2014.346.51,PLA.2015.379.11}. In the context of mesoscopic rings, rotation-induced effects have been investigated by Merlin, which
analysed the electronic contribution to the moment of inertia in metal rings under rotating effects \cite{PLA.1993.181.421}. In Ref. \cite{PRB.1996.54.1877}, Merlin and Rojo  discuss the persistent magnetic moment in rotating rings and cylinders in the mesoscopic regime. Another interesting discussion on rotation-induced effects on mesoscopic rings can be found in Ref. \cite{PLA.1995.197.444} (see also Ref. \cite{FBS.2022.63.58}), where
Vignale and Mashhoon showed that mesoscopic rings set in rotation with angular velocity $\Omega$
carry a persistent current which is a periodic function
of $\Omega$, with period $\Omega_{0}=h/2\mu S$, where $S$ is area enclosed by the ring in a plane perpendicular to $\mathbf{\Omega}$. In other words, we can state that rotation plays the role of magnetic flux. Thus, the problem of rotating rings is equivalent to the problem of non-rotating
rings in the presence of a magnetic flux. This equivalence between vector potentials and changes of the frames of references was given by Aharonov and Carmi \cite{FP.1973.3.493,FP.1974.4.75}. They provided an important analysis of the relationship between electromagnetic fields and inertial forces. The idea is to show that rotation induces a phase shift in the interference pattern of the two-slit electron diffraction experiment even in the absence of magnetic fields. A review of the results obtained by Aharonov and Carmi can be found in Ref. \cite{FP.1980.10.151}.

The rest of the paper is organized as follows. In Section \ref{sec2}, we analyze the behavior of electrons in the presence of external magnetic fields in a rotating frame. The energies are obtained, and we discuss how rotation affects them. In Section \ref{sec:Electron states in a two-dimensional quantum ring}, we consider electrons confined in a two-dimensional quantum ring. To simulate the boundaries of the quantum ring, we introduce square well potential. In this case, the energies are obtained numerically. The model has been used by Halperin \cite{PRB.1982.25.2185} as well as MacDonald and St\v{r}eda \cite{PRB.1984.29.1616} to investigate the role played by edge currents in the quantization of the Hall conductivity.
The hard wall potential model has also been used in other works. For instance by Bandos, Cantarero and  Garc\'{i}a-Crist\'{o}balto \cite{EPJB.2006.53.99} to study the excitonic Aharonov-Bohm effect in quantum rings, whereas Reimann et al. \cite{ZPBCM.1997.101.377} measured conductance oscillations in
a $2$DEG confined in a quantum dot described by hard wall potential. In Ref. \cite{PE.2022.144.115431}, Lumb et al. investigated the generated charge currents and induced magnetization in a distorted GaAs/AlGaAs quantum disk and explore the possibility of obtaining adiabatic magnetic pulses in the sub-picosecond range.
In Section \ref{Sec:NumericalResult}, we use the results from Section \ref{sec:Electron states in a two-dimensional quantum ring} to numerically study the electronic states of the ring, Fermi energy and magnetization. A detailed study of physical properties such as magnetization is difficult because of the complexity of the spectrum shape. However, by using some approximations, it was possible to obtain an expression for the magnetization, whose results are consistent with the literature when the ring is in an inertial frame of reference. When we added the rotation, we found that the analytical model was not viable. In this case, we consider another approach to obtain the magnetization and the effects of rotation on it. Our results are summarized in Section \ref{Sec:conclusoes}.

\section{Landau levels}
\label{sec2}

In this section, we present the physical environment under which we shall consider the motion of the electrons. We first write the Hamiltonian for a particle at rest with respect to the rotating, followed by the insertion of the vector potential associated with a uniform magnetic field which allow us to obtain analytical solutions of the Schr\"{o}dinger equation. According to Ref. \cite{Book_Rotating_Frames} the inclusion of rotating effects in nonrelativistic quantum mechanics can be accomplished through minimal substitution
\begin{equation}
p^{\tau }\rightarrow p^{\tau }-\mu\mathcal{A}^{\tau },
\end{equation}
where $\mathcal{A}^{\tau}$ is the gauge field for the rotating frame, with $\tau=0,1,2,3$ and $\mu$ is the mass of the particle. The gauge field $\mathcal{A}^{\tau}$ is defined by
\begin{equation}
\mathcal{A}^{\tau }=\left( \mathcal{A}_{0},\boldsymbol{\mathcal{A}}\right)=\left( -\frac{1}{2}V^{2},\mathbf{V}\right).
\label{fieldA}
\end{equation}
where $\mathbf{V}$ is the velocity. In the the rotating frame, the velocity $\mathbf{V}$ is given by ${\mathbf{\Omega }\times \mathbf{r}}$, where $\mathbf{\Omega}$ is the angular velocity. The model also includes the interaction of the electron with a uniform magnetic field. This interaction is included in the Hamiltonian of the system through the prescription  \begin{equation}
p^{\tau}\rightarrow p^{\tau}-eA^{\tau},
\end{equation}
where $e<0$ is the charge of the particle and $A^{\tau}$ is the four-potential. Here, the four-potential of the electromagnetic field $A^{\tau}$ has only the spatial component, $A^{\tau}=(0, \mathbf{A})$, where $ \mathbf{A}$ is the vector potential.

We assume that the magnetic field is normal to the plane of electronic motion. In polar coordinates, the vector potential that describes this configuration of magnetic field is given by symmetric gauge
\begin{equation}
\mathbf{A}=\frac{Br}{2}\mathbf{\hat{\varphi}}.
\label{calibre}
\end{equation}

In this way, the Hamiltonian for a particle at rest with respect to the rotating frame is written as
\begin{equation}
\mathcal{H}=\frac{1}{2\mu} \left(\mathbf{p}-e\mathbf{A}-\mu\mathbf{\Omega }\times \mathbf{%
r}\right)^{2}-\frac{1}{2}\mu \left(\mathbf{\Omega }\times \mathbf{r}\right)^{2}.
\label{schrodingerA}
\end{equation}

We also assume that the motion occurs in the $xy$ plane and that in the rotating frame the angular velocity has only the z-component, i.e., $\mathbf{\Omega}=\Omega \,\mathbf{\hat{z}}$, with $\Omega \geq 0$. In this case, the Schr\"odinger equation (in polar coordinates) takes the form
\begin{align}
&\Bigg[-\frac{\hbar^{2}}{2\mu }\frac{1}{r}\frac{\partial }{\partial r}
\left( r\frac{\partial }{\partial r}\right)-\frac{\hbar^{2}}{2\mu} \Bigg( \frac{1}{r^{2}} \frac{\partial^{2}}{\partial \varphi^{2} }-i\frac{\mu \omega_{2}}{\hbar} \frac{\partial}{\partial \varphi }\notag \\
&-\frac{\mu^{2}\omega_{1}^{2}}{4\hbar^{2}}r^{2}\Bigg)\Bigg]\psi_{n,m}=E_{n,m}\psi_{n,m},
\label{Eq.S.completa}
\end{align}
where
\begin{equation}
\omega_{1}=\sqrt{\omega_{c}^{2}+4\omega_{c}\Omega},
\;\;\;\;\
\omega_{2}=\omega_{c}+2\Omega, \label{om}
\end{equation}
are the effective frequencies, and $\omega _{c}=eB/\mu$ is the cyclotron
frequency.

By using wave functions in the form
\begin{equation}
\psi_{n,m} \left( r,\varphi \right) =e^{im\varphi}f_{n,m}(r),
\label{wf}
\end{equation}%
with $m=0,\pm 1,\pm 2,\ldots $, and substituting in Eq. (\ref{Eq.S.completa}), we obtain the radial equation
\begin{equation}
\left[ -\frac{\hbar ^{2}}{2\mu }\frac{1}{r}\frac{d}{dr}\left( r\frac{d}{dr}%
\right) +V_{eff}\left( r\right) \right] f_{n,m}=E_{n,m}^{'}f_{n,m},
\label{em.6}
\end{equation}%
where
\begin{equation}
V_{eff}\left( r\right) =\frac{\hbar^{2}}{2\mu}\left( \frac{m^{2}}{r^{2}%
}+\frac{\mu^2\omega_{1}^{2}
}{4\hbar^{2}  }r^{2}-\frac{|m|\mu\omega_{1}}{\hbar}\right)
\label{pot.efe.}
\end{equation}
is the effective potential, and $E_{n,m}^{\prime }=E_{n,m}-V_{min}$,
where $V_{min}$ is the minimal energy of the states, given by
\begin{equation}
V_{min}=\frac{1}{2}\hbar\left(|m|\omega_{1}-m\omega_{2}\right),
\label{Vmin}
\end{equation}
The effective potential given by Eq. (\ref{pot.efe.}) has a minimum at
\begin{equation}
r_{m}=\lambda_{1}\sqrt{2\left\vert m\right\vert },
\label{raio.m}
\end{equation}
where $\lambda_{1}$ is the effective magnetic length, given by
\begin{equation}
\lambda_{1}=\sqrt{\frac{\hbar}{\mu \omega_{1}}}.
\label{Eq:Com.Mag.Efet.}
\end{equation}
Equation (\ref{raio.m}) determines the radial position of the states, is that, is a measure of the
relative position of the envelope function associated with a state of momentum angular $m$. Furthermore,
for $r\rightarrow r_{m}$, the effective potential has the simple parabolic form
\begin{equation}
V_{par}\left( r\right) = \frac{1}{2}\mu \omega _{1}^{2}\left( r-r_{m}\right)
^{2}.
\label{Eq:pot.parab}
\end{equation}

The solution of Eq.~(\ref{em.6}) is well known and is given in terms of the confluent hypergeometric function of the first kind \cite{Book.1981.Landau}. The normalized wave eigenfunctions read
\begin{align}
\psi_{n,m} \left( r, \varphi \right)=& \frac{1}{\lambda_{1}}\sqrt{\frac{\Gamma
(\left| m\right|+n+1)}{2^{\left| m\right|+1} n!\,\Gamma (\left|
m\right|+1)^2\pi}}\,e^{im\varphi} e^{ -\frac{r^2}{4\lambda_{1}^2}}\notag \\
&\times\left(\frac{r}{\lambda_{1}}\right)^{\left| m \right|}{_{1}F_{1}}%
\left(-n,\left| m \right|+1, \frac{r^2}{2\lambda_{1}^2}\right),
\label{funcao.onda.1}
\end{align}
where $n=0,1,2,\dots$, $\Gamma \left( x \right)$ is the gamma function, and $_{1}F_{1} \left( a, b; x \right)$ is the confluent hypergeometric function of the first kind \cite{abramo}.
The corresponding energy eigenvalues are
\begin{equation}
E_{n,m}=\left(n+\frac{1}{2}\right) \hbar \omega_{1} + V_{min},
\label{energy.1}
\end{equation}
where $V_{min}$ is given by Eq. (\ref{Vmin}).

\begin{figure}[!b]
\includegraphics[scale=0.4]{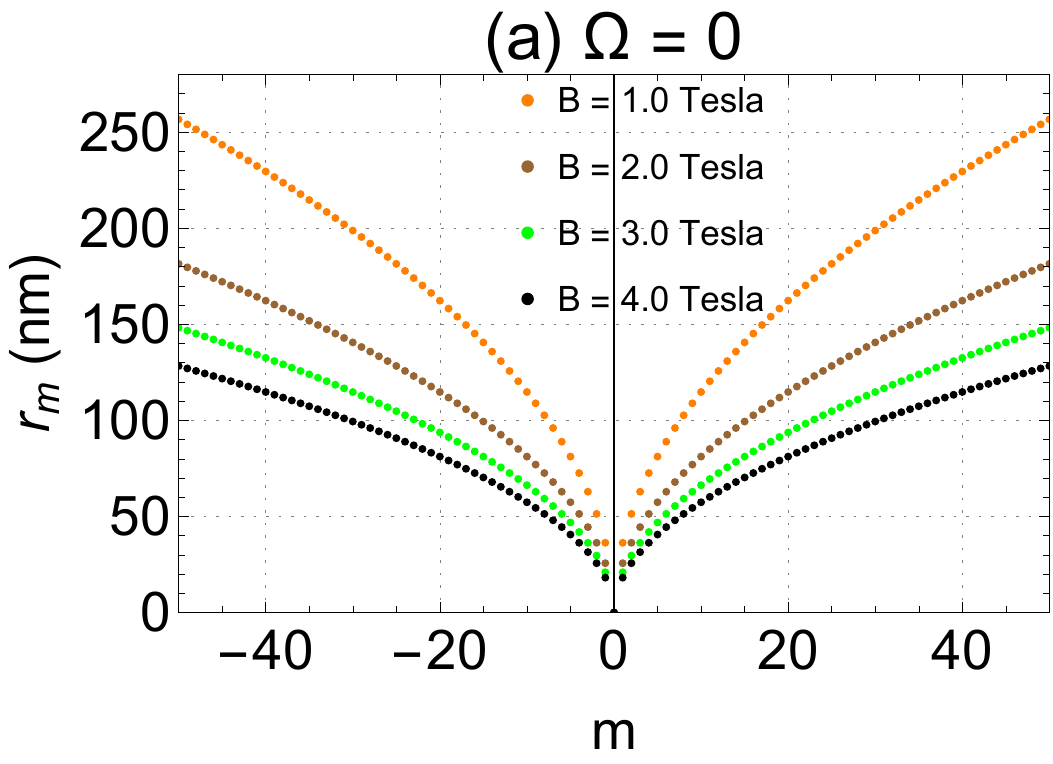}
\includegraphics[scale=0.4]{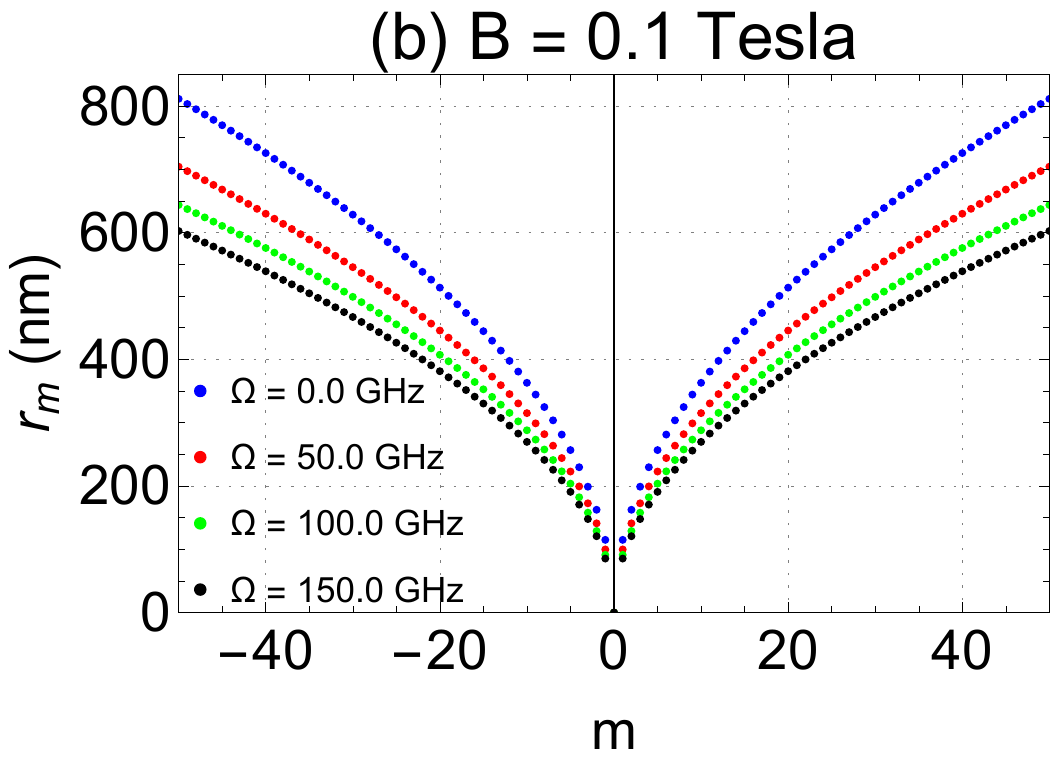}
\caption{Sketch of $r_{m}$ as a function of $m$. In (a), we use $\Omega=0$ Hz and some values of $B$ while in (b) the magnetic field is fixed at $B=0.1$ Tesla and some values of $\Omega$.}
\label{Fig:raiodosestados}
\end{figure}

The radius of a state given by Eq. (\ref{raio.m}) depends on both the
angular number $m$ and magnetic field strength as well as angular velocity. States with quantum number $m$ and $-m$ have the same radius. The radius $r_{m}$ shrink as the strength of the magnetic field or the angular velocity is increased. These results are displayed in Fig. \ref{Fig:raiodosestados}. The flux enclosed between consecutive radii is given by
\begin{equation}
\pi B \left( r_{m+1}^{2}-r_{m}^{2} \right) = \phi_{0}\sqrt{\frac{\omega_{c}}{\omega_{c}+4\Omega}}.
\end{equation}

It is interesting to investigate the rotating effects in the energy eigenvalues, given by Eq. (\ref{energy.1}). When $\Omega=0$, Eq. (\ref{Vmin}) shown that states with $m \geq 0$ have zero minimum potential energy. In that
case, as shown in Eq. (\ref{energy.1}), the energy does not depend on angular number $m$. It follows from this result that the states with $m \geq 0$ with the same quantum number $n$ have the same energy.  Then, there is the formation of infinitely degenerate energy levels, which are called Landau levels. The energy gap between the consecutive Landau levels corresponds to $\hbar \omega_{c}$. States with $m<0$ belong to Landau levels of quantum number $n+|m|$.  In fact, the states with angular number $m<0$ have a minimum potential energy given by $|m|\hbar \omega_{c}$. On the other hand, when $\Omega>0$ all states have a minimum potential energy and are proportional to a combination of both the frequencies $\omega_{1}$ and $\omega_{2}$.  As the first term of Eq. (\ref{energy.1}) is proportional to $\omega_{1}$, then the Landau degeneracy is lifted. In addition, the spacing between consecutive energy levels is given by $\hbar \omega_{1}$. Therefore, the energy separation between the adjacent levels increases as the strength of the magnetic field or the angular velocity is increased.

\begin{figure}[!h]
\includegraphics[scale=0.39]{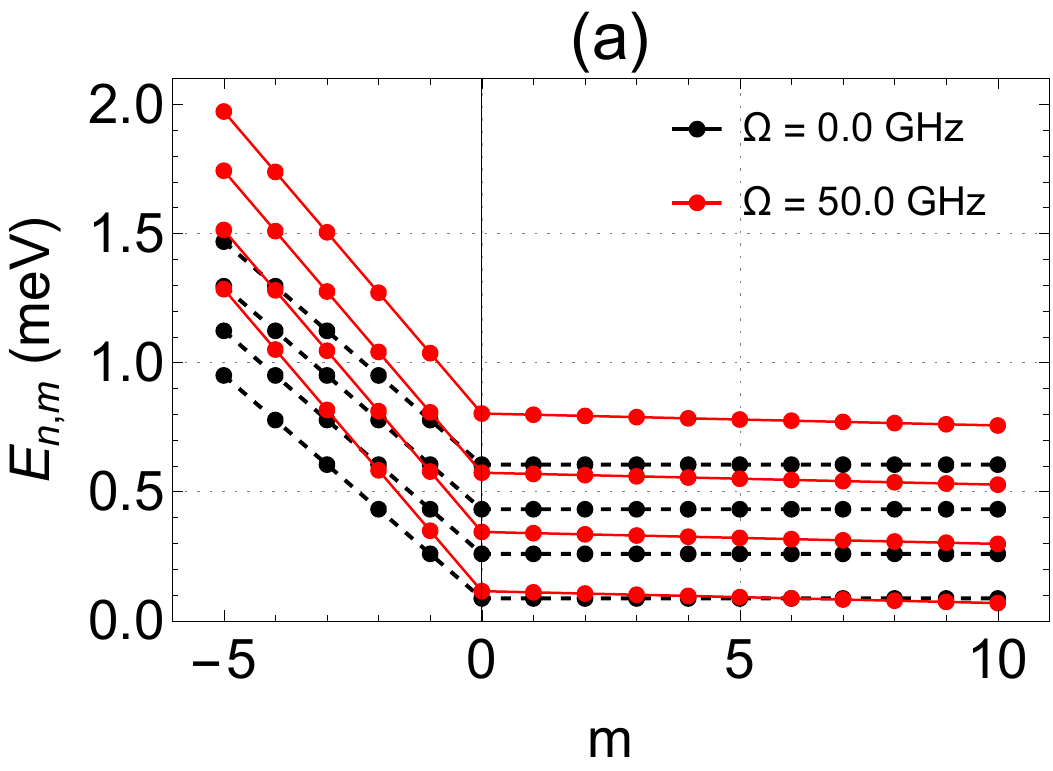}
\includegraphics[scale=0.39]{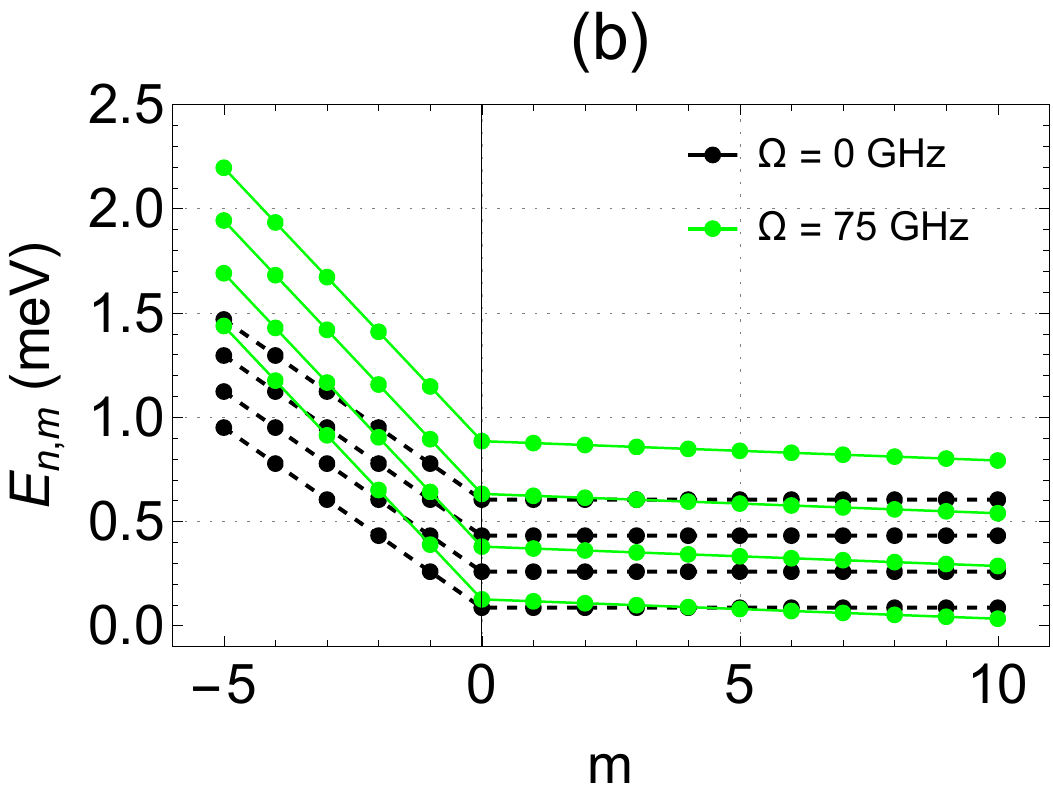}
\includegraphics[scale=0.39]{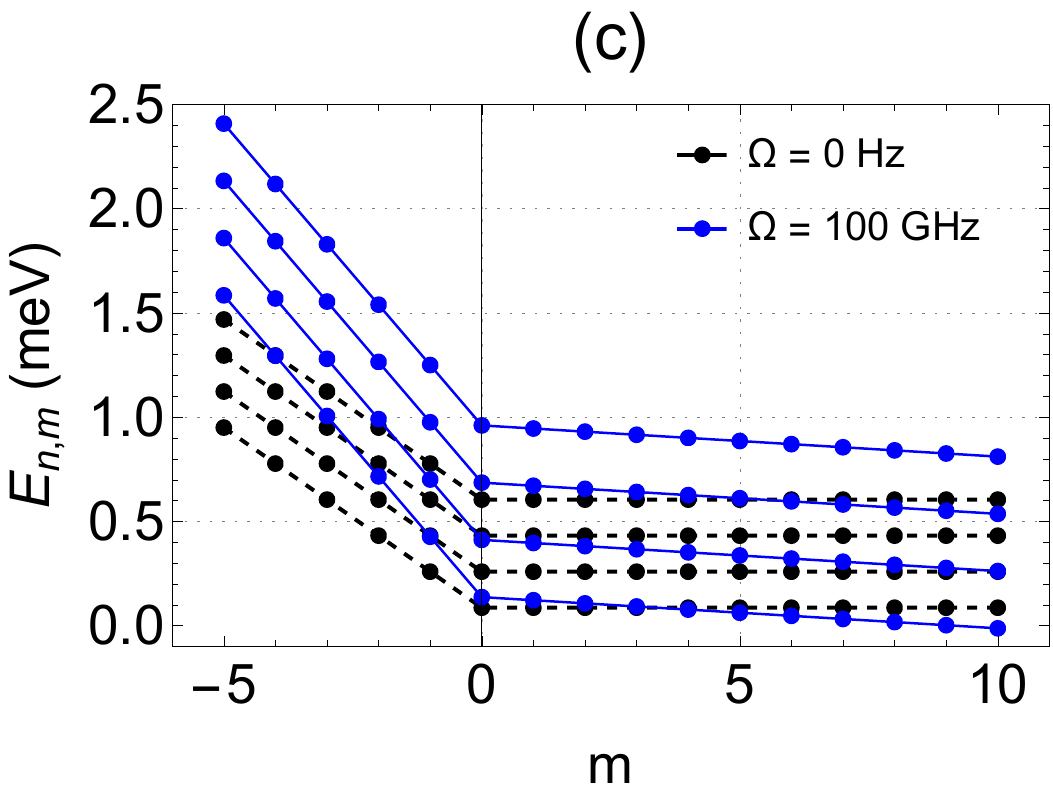}
\includegraphics[scale=0.39]{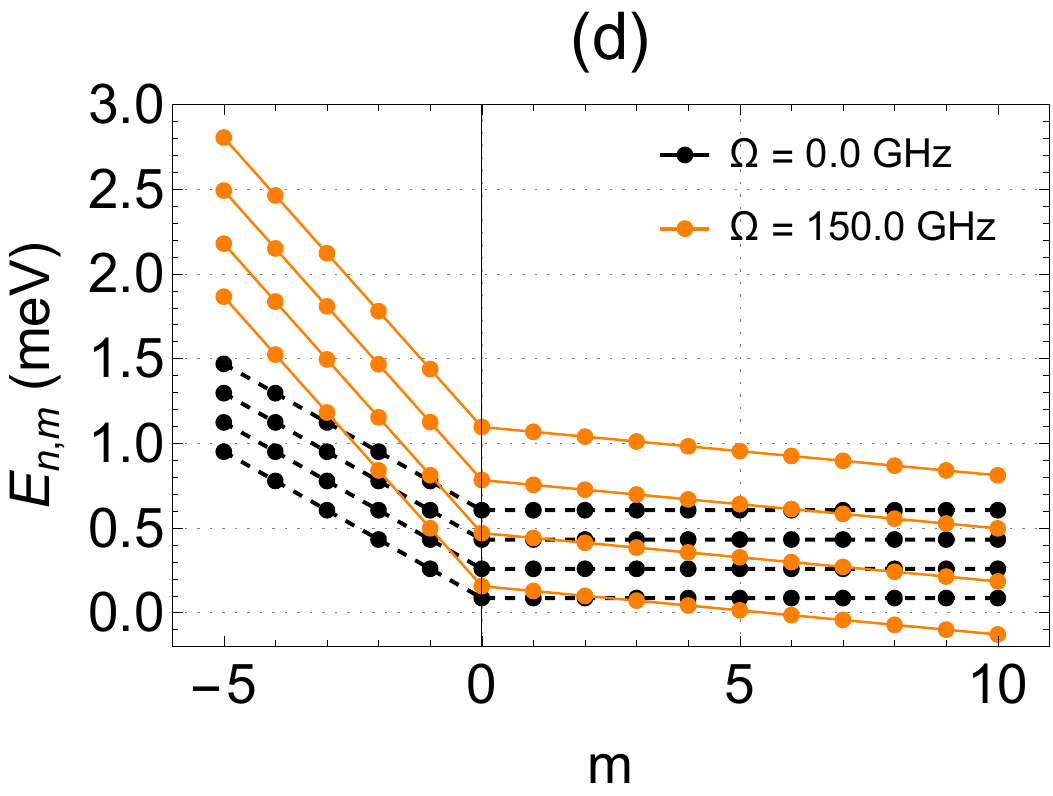}
\caption{The first four Landau levels as a function of the quantum number $m$ for $B=0.1$ Tesla. We can clearly see the effects of the minimum potential energy $E_{m}$ on the states with $m>0$ and $m<0$ for different values of $\Omega$. Landau levels with $\Omega=0$ are displayed together with the levels where (a) $\Omega=50$ GHz, (b) $\Omega=75$ GHz, (c) $\Omega=100$ GHz and (d) $\Omega=150$ GHz.}
\label{energylandau}
\end{figure}

In Fig. \ref{energylandau}, we show the first four Landau levels for the case  $B=0.1$ Tesla. The dots and continuous lines correspond to the states and levels, respectively. We can see the effect of the minimum potential on the states: if $\Omega=0$, then the states with $m \geq 0$ are degenerate, while the states with $m<0$ belong to higher Landau levels. In other hand, the angular velocity increase the energy separation between the adjacent levels. The slope of the Landau levels for states with $m>0$ is given by $\hbar(\omega_{1}-\omega_{2})/2$, that is, the minimum potential for these states is negative. In fact, we can show that $\omega_{1}<\omega_{2}$. In the other hand, for states with $m<0$, the slope is given by $\hbar(\omega_{1}+\omega_{2})/2$. If $\Omega=0$ in Eq. (\ref{om}), we find $\omega_{1}=\omega_{2}=\omega_{c}$. So, the rotation increases the slope of the Landau levels for states with $m<0$. Obviously, the minimum potential of the states with $m<0$ is always non-zero and positive for any value of $\Omega \geq 0$.
We verify that the Landau-levels degeneracy has been lifted by the rotation, as stated above. Furthermore, Fig \ref{energylandau} shows also that as angular velocity increases, the rotating effects on the Landau states increase.

\section{Edge states in a two-dimensional quantum ring}

\label{sec:Electron states in a two-dimensional quantum ring}

\begin{figure}[!t]
\centering
\includegraphics[scale=0.55]{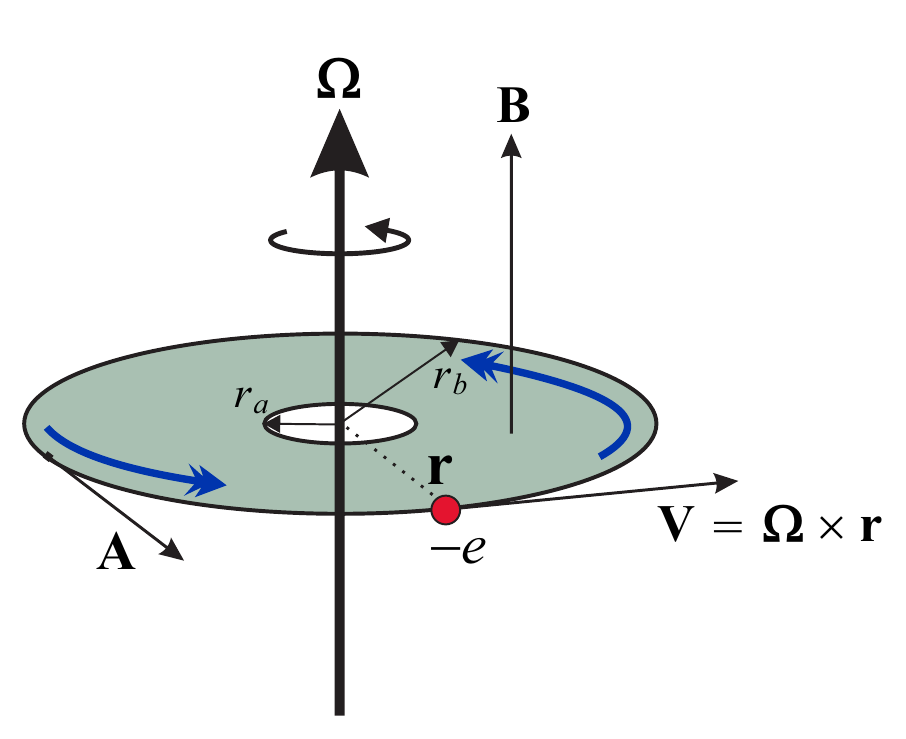}
\caption{Idealization of a 2DEG confined to a rotating disk in a uniform magnetic field and Aharonov-Bohm flux. The electrons are constrained to the strip delimited by the radii $r_{a}$ and $r_{b}$.}
\label{model}
\end{figure}

In this section, we study the model of a 2DEG confined to a mesoscopic ring of radii inner and outer, $r_{a}$ and $r_{b}$, respectively. The ring is in a rotating frame and subjected to uniform magnetic field $B$ perpendicular to its plane as shown in Fig. \ref{model}.

The confining potential used to model ring is
assumed to be of the following form,
\begin{equation}
V\left( r\right) =\Bigg\{
\begin{array}{ll}
0, & \mbox{if} \quad r_{a}\leq r\leq r_{b} \\
\infty, & \quad \text{otherwise}
\end{array}.
\label{Eq:HardWall}
\end{equation}
If the states occupy the  region next middle of the ring and away from the edges, the confining potential $V(r)$ does not influence the behavior of the states. Thus, it is evident that the equation for the harmonic oscillator given by Eq. (\ref{em.6}) describes the radial motion of the electron. Consequently, wavefunctions and eingevalues are given by Eqs. (\ref{funcao.onda.1}) and (\ref{energy.1}), respectively.

For states localized near the edges of the quantum
ring, however, we must take into account the effects of the confining potential $V\left(r\right)$. In fact, the edges of the sample are modeled by hard walls as shown in Eq. (\ref{Eq:HardWall}). We can analyze the electronic profile at the inner edge, and ignore the effect of the outer edge. Subsequently, we must seek solutions that vanish as $r-r_{m}\rightarrow \infty$. The wave functions have the form
\begin{equation}
\chi_{\nu,m}\left( r,\varphi \right) =e^{im\varphi}f_{\nu,m}(r),
\end{equation}
where $f_{\nu,m}(r)$ satisfies the radial differential equation  \cite{PRB.1982.25.2185,PRB.1984.29.1616}
\begin{equation}
\left[ -\frac{\hbar ^{2}}{2\mu }\frac{d^{2}}{dr^{2}}+V_{par}+V(r) \right] f_{\nu,m}=E_{\nu,m}^{'}f_{\nu,m}.
\label{eq.sch}
\end{equation}
Here $E_{\nu,m}^{'}=E_{\nu,m}-V_{min}$, where $V_{min}$ is given by Eq. (\ref{Vmin}), $V_{par}$ is the parabolic potential given in Eq. (\ref{Eq:pot.parab}), and $V\left(r\right)$ is the  confining potential, defined by Eq. (\ref{Eq:HardWall}).

The solution of Eq. (\ref{eq.sch}) satisfying the condition $f_{\nu,m}(r) \rightarrow 0$ as $r-r_{m} \rightarrow \infty$ is given by~\cite{PRB.1984.29.1616,Book.merzbacher.1998}
\begin{align}
f_{\nu,m}\left(r \right)& =e^{-\frac{1}{2}\left(\frac{r -r_{m}}{\lambda_{1}}\right)^{2}}  \Bigg[c_{1}\mathrm{M}\left( -\frac{\nu }{2},\frac{1}{2}
;\left(\frac{r -r_{m}}{\lambda_{1}}\right)^{2}
\right) \notag \\
&+ c_{2}\left(\frac{r -r_{m}}{\lambda_{1}}\right)  \mathrm{M}\left( \frac{1}{2}-\frac{\nu }{2},\frac{3}{2};\left(\frac{r -r_{m}}{\lambda_{1}}\right)^{2}\right) \Bigg],
\label{weber}
\end{align}
where $c_{1}$ and $c_{2}$  are constant, and $\nu$ correspond to the zeros of $f_{\nu,m}(r)$. Here, $\nu$ is not only restricted to take integer values, namely, $\nu$ can take any positive real values.

The energy spectrum of the system is given by
\begin{equation}
E_{\nu ,m}=\left( \nu +\frac{1}{2}\right) \hbar \omega _{1}+V_{min},
\label{energia.2}
\end{equation}
where $\omega_{1}$ and $V_{min}$ are given by Eqs. (\ref{om}) and (\ref{Vmin}), respectively.
The requirement that the wave function $f_{\nu,m}(r) \rightarrow 0$ as $r \rightarrow \infty$ leads to the following result \cite{PRB.1984.29.1616}: \begin{equation}
\frac{c_{2}}{c_{1}}=2\frac{\Gamma(1+\frac{\nu}{2})}{\Gamma(\frac{1}{2}+\frac{\nu}{2})}\tan(\frac{\pi \nu}{2}).
\label{eq.transcendental}
\end{equation}
Furthermore, we shall impose that the wave function vanishes at $r=r_{a}$. Combining this result with the above equation, we compute the values of $n$ numerically.

For states located far from the  edges, i.e., those with large values of  $r_{a}-r_{m}$ compared to the magnetic length $\lambda_{1}$, Eq. (\ref{eq.transcendental}) provides
\begin{equation}
\sin{\pi \nu}=0,
\label{zeros.2}
\end{equation}
which means that $\nu = n = 0,1,2\ldots$. In this way, we recover the result obtained at the beginning of this Section, namely that states located inside the ring, but far from the edges, where the confinement potential does not influence the electronic states, the energy eigenvalues are given by Eq. (\ref{energy.1}). In addition, if $\Omega=0$, we recover the results obtained by Halperin \cite{PRB.1982.25.2185} and MacDonald and  St\v{r}eda \cite{PRB.1984.29.1616}.

\begin{figure}[b]
\centering
\includegraphics[scale=0.39]{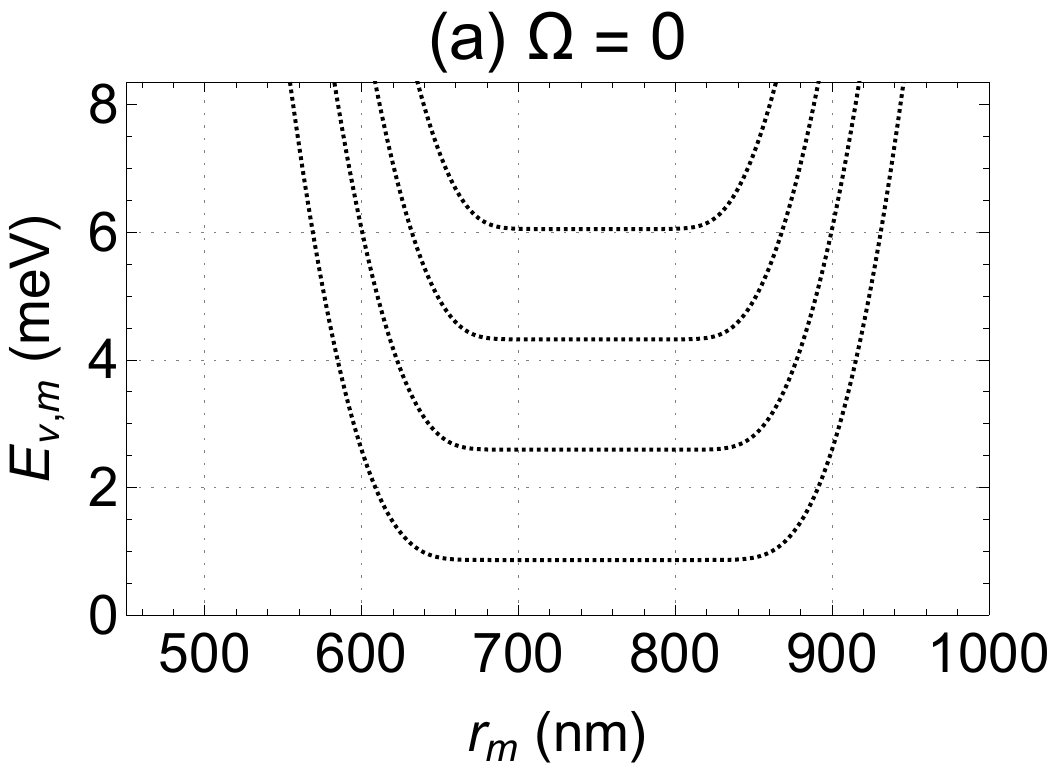}
\includegraphics[scale=0.39]{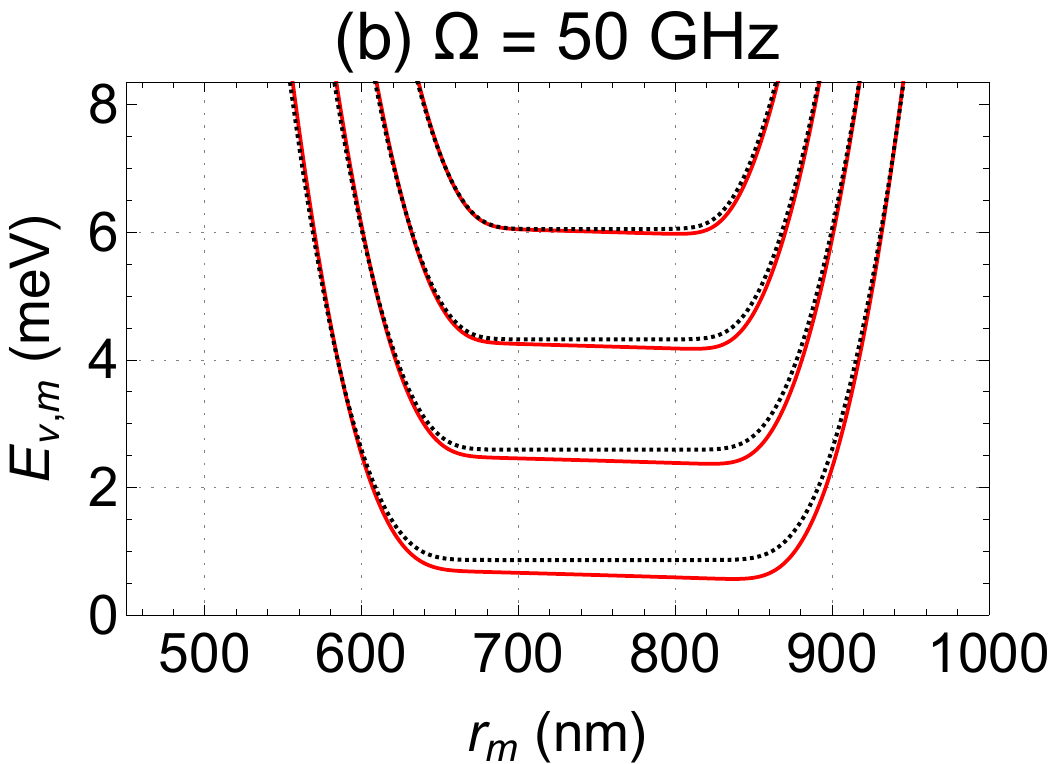}
\includegraphics[scale=0.39]{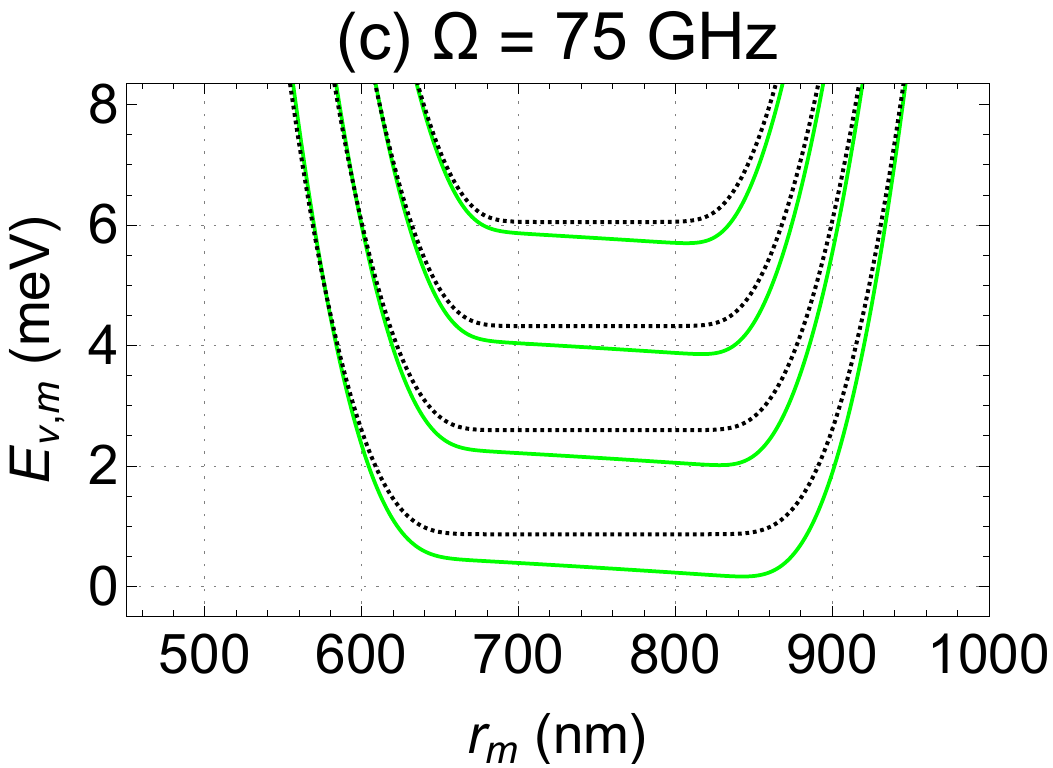}
\includegraphics[scale=0.39]{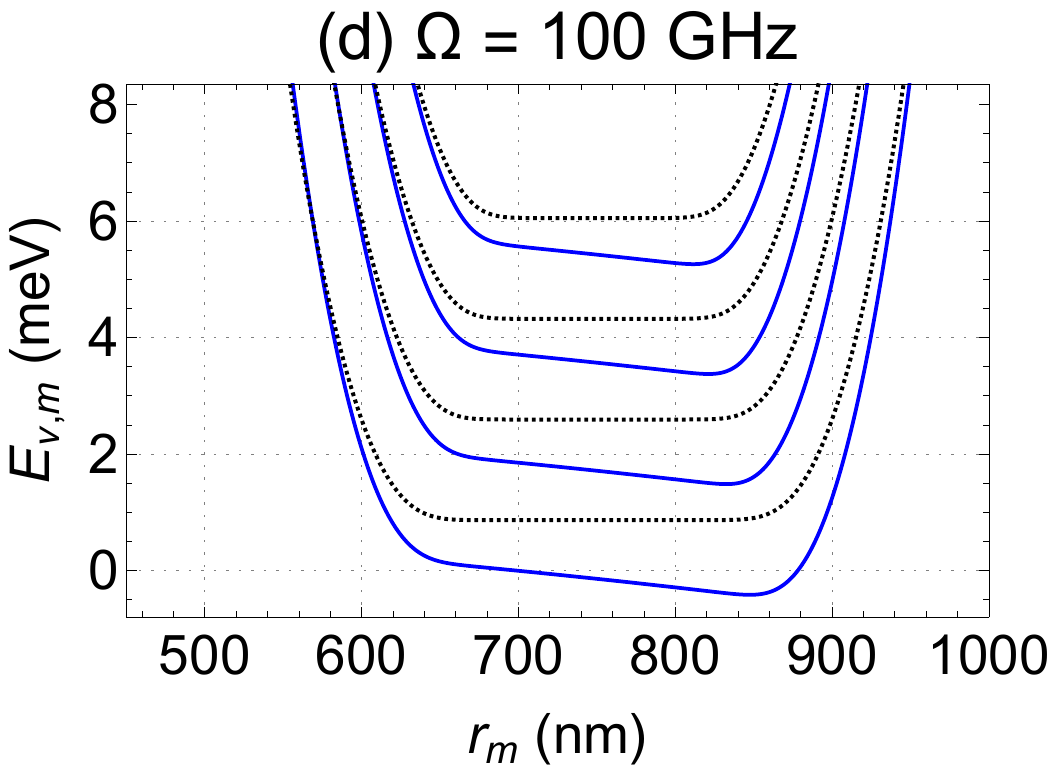}
\caption{Energy levels of a 2D quantum ring (Eq. (\ref{energia.2})) as a function of the radial $r_{m}$. When $\Omega>0$, a slope appears in the subbands.}
\label{figureenergy}
\end{figure}

\section{Numerical results}

\label{Sec:NumericalResult}

In this section, we perform a numerical approach to the results found in Sec. \ref{sec:Electron states in a two-dimensional quantum ring}. For our purposes, we shall investigate the energy spectrum, the Fermi energy and the magnetization of a 2D quantum ring. The sample consists of a quantum ring with inner and outer radii given by $r_{a}=600$ nm and $r_{b}=900$ nm, respectively. The heterostructure is made of GaAs and the effective mass of the electron is $\mu=0.067 \mu_{e}$, where $\mu_{e}$ is the electron mass. In the numerical analyse of both Fermi energy and the magnetization as a function of the magnetic field, we consider a sample with $N=1100$ spinless electrons.

\subsection{Energy}

In Fig. \ref{figureenergy}, we plot the energy $E_{\nu,m}$ (Eq. (\ref{energia.2})) as a function of $r_{m}$ (Eq. (\ref{raio.m})) for $B=1.0$ Tesla and different angular velocities $\Omega$.
We can observe that the behavior of the electronic states changes according to the position they occupy inside the ring. Near the middle of the ring, the states do not see the confining potential. For this case, the values of $\nu$ are computed from Eq. (\ref{zeros.2}), and consequently the energies are simply given by Eq. (\ref{energy.1}).
We have already discussed the results of this configuration in Sec. \ref{sec2}. Such states are known as bulk states. In contrast, electronic states centered near the to ring edges have a strong dependence on quantum number $m$. In fact, as we see in Fig. \ref{figureenergy}, the energy $E_{\nu,m}$ increases monotonically
as $r_{m}$ approaches one of the edges. These states are commonly referred as edge states.
Is evident the slope of subbands when $\Omega>0$. In effect, the minimum potential energy (\ref{Vmin}) is negative for states with $m>0$ (see Sec. \ref{sec2}).

\begin{figure}[t!]
\centering
\includegraphics[scale=0.25]{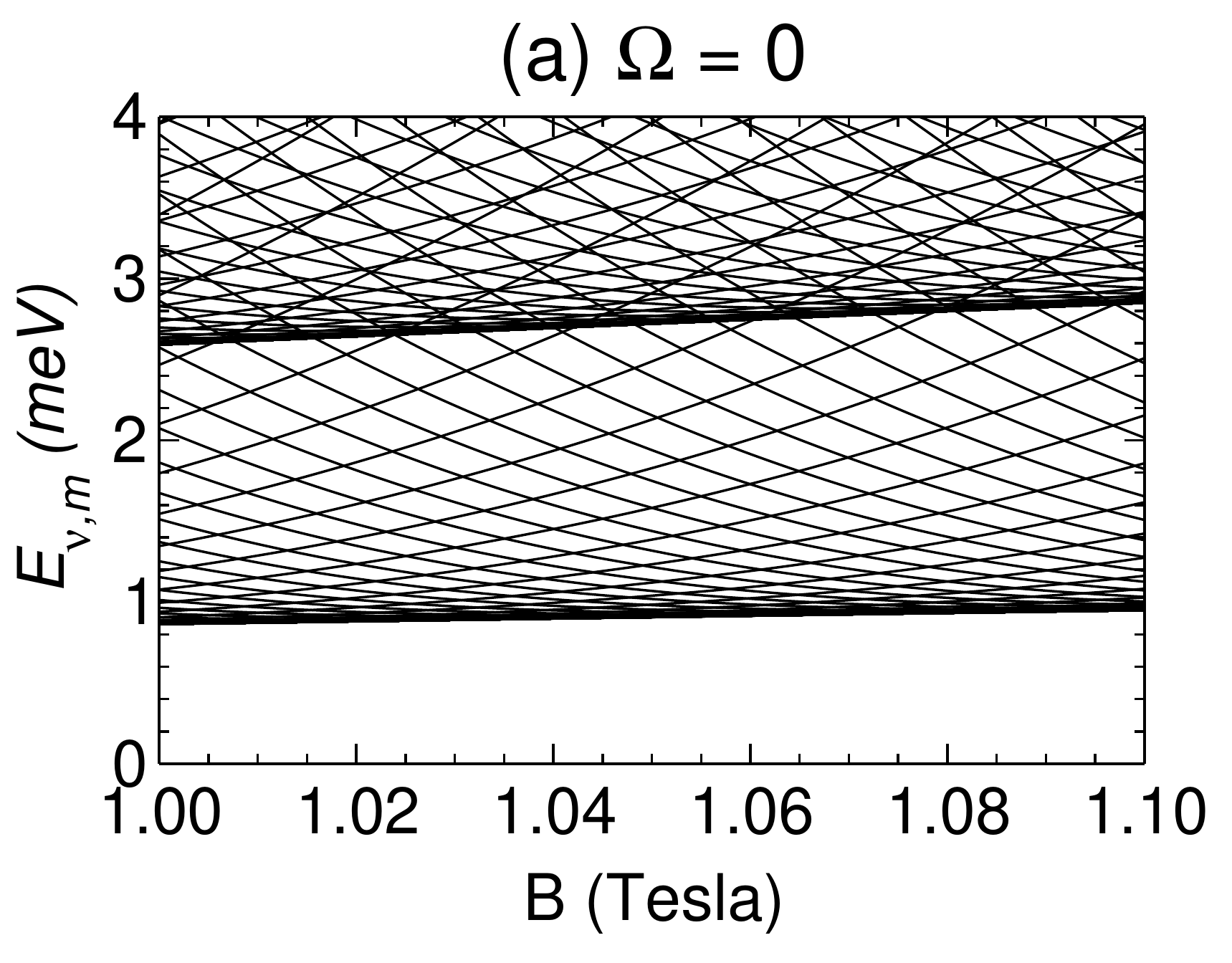}
\includegraphics[scale=0.25]{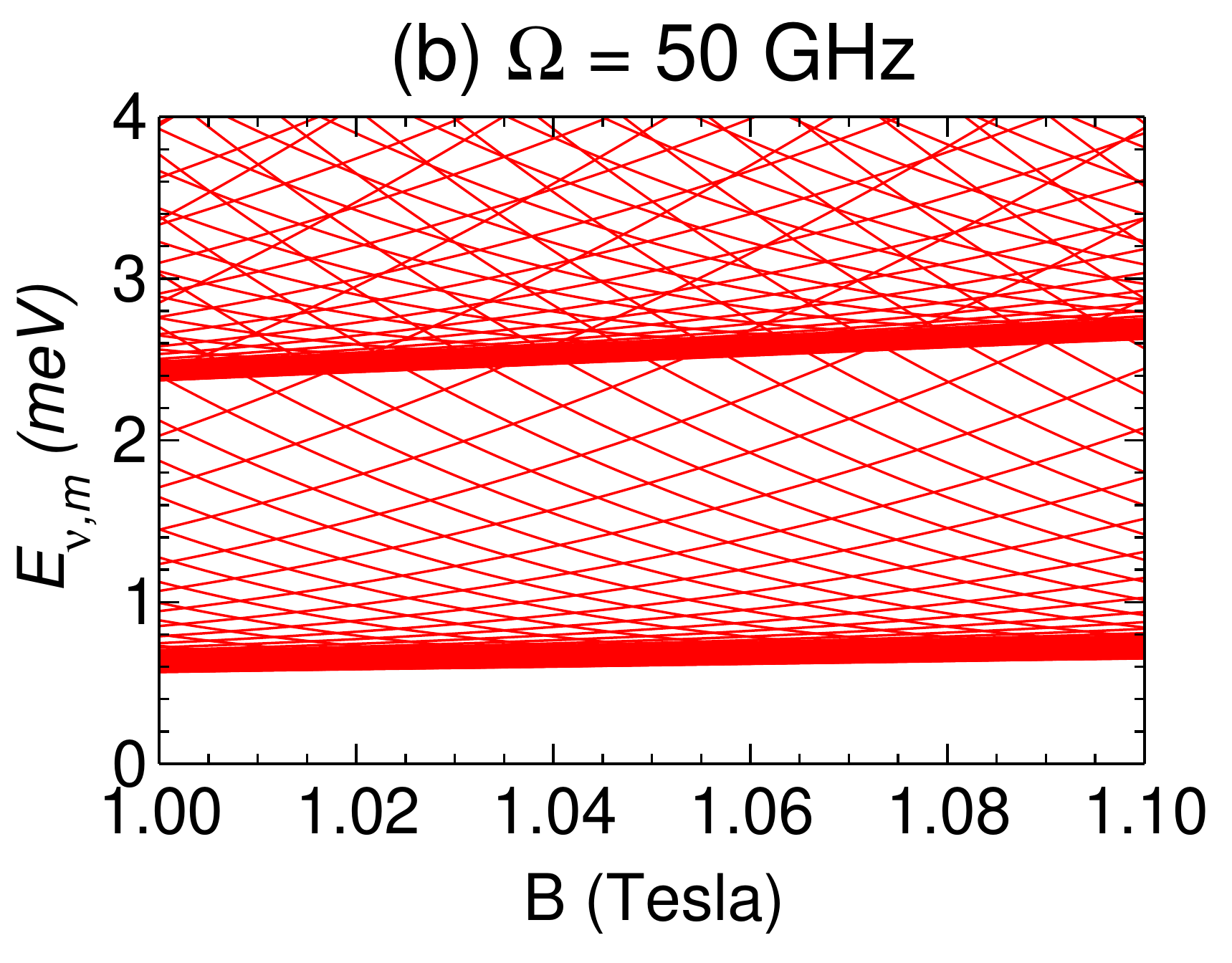}
\includegraphics[scale=0.25]{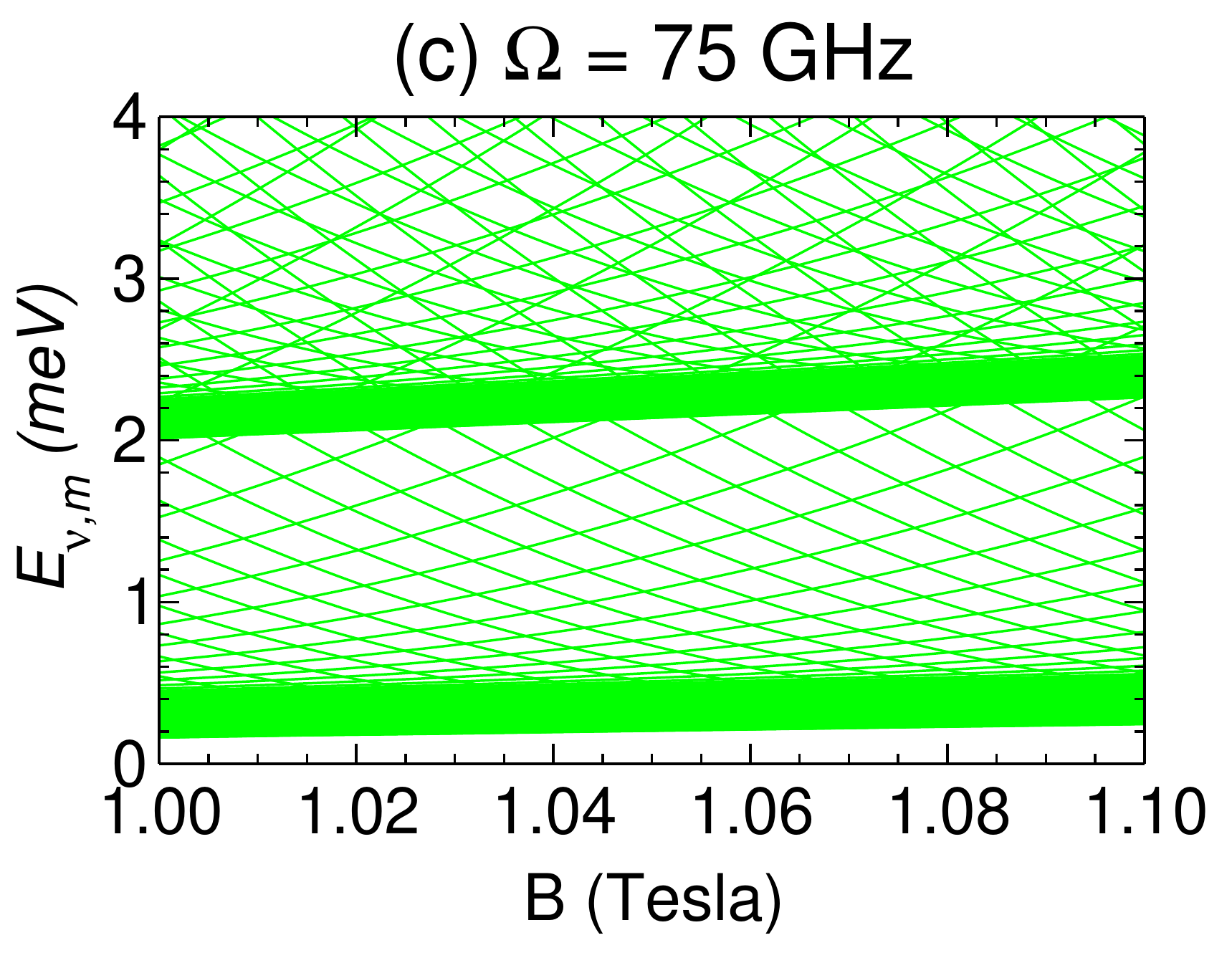}
\includegraphics[scale=0.25]{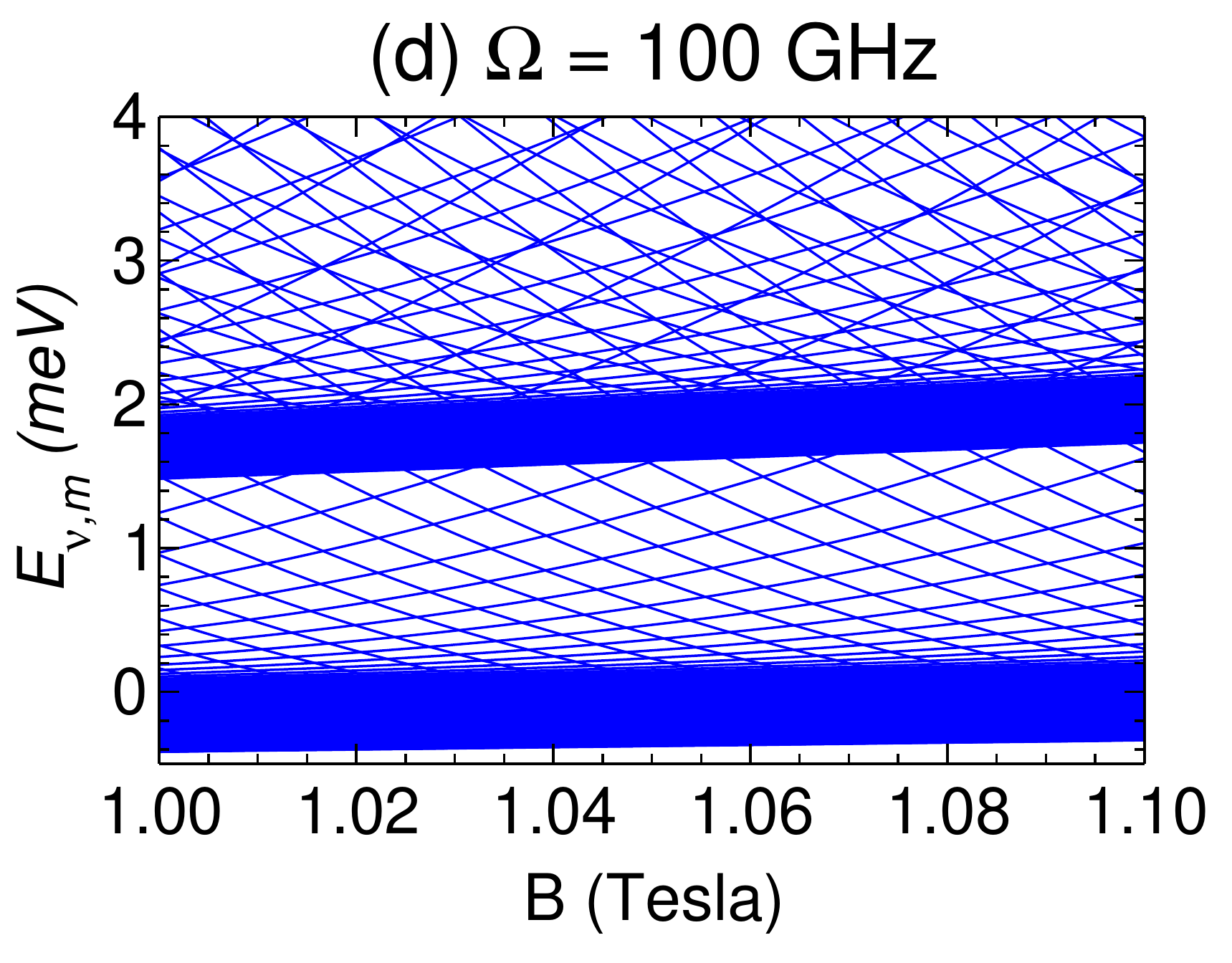}
\caption{Energy levels of a 2D quantum ring (Eq. (\ref{energia.2})) as a function of the magnetic field.}
\label{Fig:EnmxB}
\end{figure}

\begin{figure}[t!]
\centering
\includegraphics[scale=0.30]{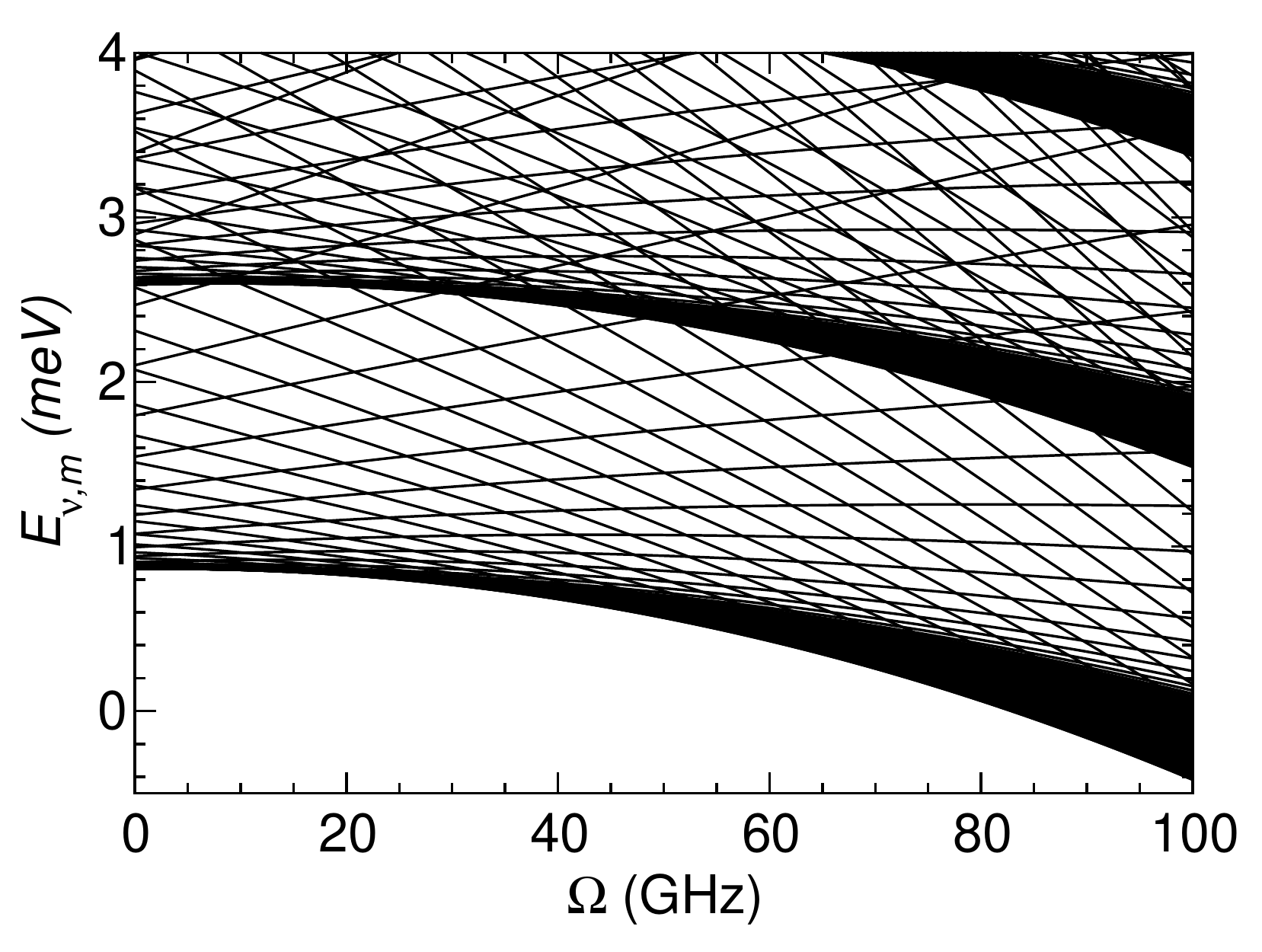}
\caption{Energy levels of a 2D quantum ring (Eq. (\ref{energia.2})) as a function of $\Omega$. We use $B=1.0$ Tesla.}
\label{Fig:EnmxW}
\end{figure}

In Fig. (\ref{Fig:EnmxB}), we show the evolution of the electronic states as a function of the magnetic field for the two lowest sub-bands. States belonging
to outer edge reduce their
energy as $B$ is increased, while those states belonging to inner edge are running upwards in energy. Figure \ref{Fig:EnmxB}(a) shows that the states of the 2D quantum ring converge
into degenerate Landau levels at the bottom of the subbands. On the other hand, Figs. \ref{Fig:EnmxB}(b)-(d) reveal that there is no convergence, instead there is a high concentration of states
in region near the bottom of the subbands.
The reason for this behavior is that the degeneracy at the bottom of the subbands (Landau levels) is removed due to rotation effects.
These results can also be seen in Fig. \ref{Fig:EnmxW}, where we evaluate the energy as a function of $\Omega$ for $B=1.0$ Tesla.
In fact, we can verify the formation of Landau levels at the bottom of the subbands for small values of $\Omega$. By increasing $\Omega$, the lifted of the degeneracy of the Landau levels becomes more evident. These results agree with the analysis of the Figs. \ref{figureenergy} and \ref{Fig:EnmxB}. Note the different effects of the magnetic field and rotation at the bottom of the subbands in Figs \ref{Fig:EnmxB} and \ref{Fig:EnmxW}, respectively. Indeed, the bottom of the subbands increases as $B$ increases. On the other hand, the bottom of the subbands decreases as $\Omega$ increases because of the influence of the minimum potential.

\begin{figure}[!b]
\includegraphics[scale=0.4]{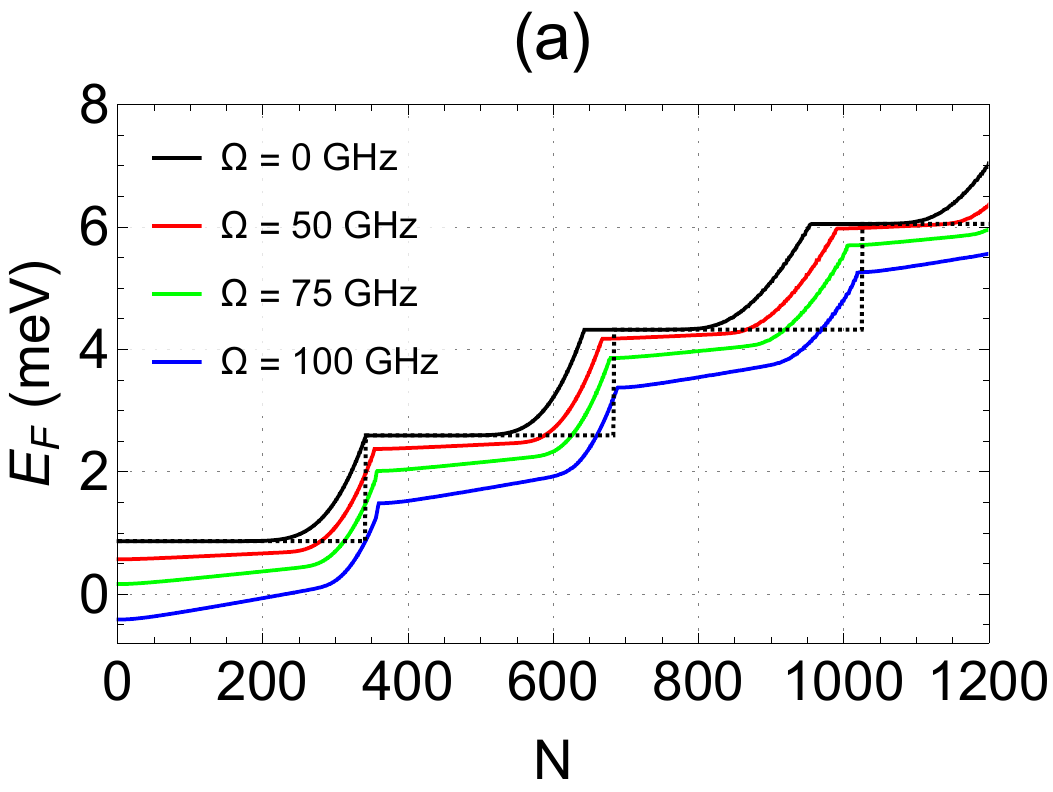}
\includegraphics[scale=0.4]{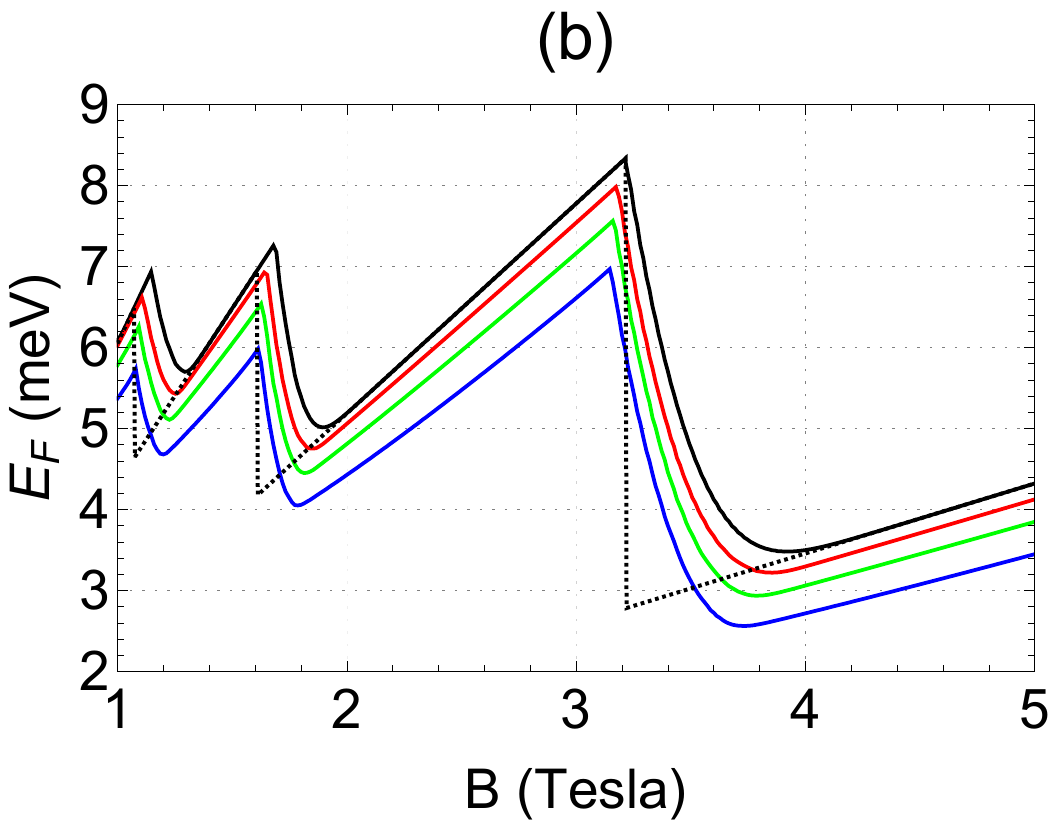}
\caption{Fermi energy of a 2D ring as a function of number of electrons (panel (a)) and the magnetic field (panel (b)).}
\label{FIG:fermi.N}
\end{figure}

\subsection{Fermi energy}

The results described above have direct physical implications in the behavior of the Fermi energy of the system. The Fermi energy is computed self-consistently through the expression
\begin{equation}
N=\sum_{n,m}\theta \left(E_{F}-E_{\nu,m}
\right),
\label{Eq:Fermi}
\end{equation}
where $\theta(x)$ is step function.

In Fig. \ref{FIG:fermi.N}, we plot the Fermi energy as a function of the the magnetic field and the number of electrons. In Figs. \ref{FIG:fermi.N}(a) and \ref{FIG:fermi.N}(b), the black dashed line correspond to the usual Fermi energy of a quantum ring without edge states ($2$DEG bulk). When $\Omega=0$, the degenerate Landau levels in the subbands lead to the formation of plateaus (Fig. \ref{FIG:fermi.N}(a)). In other words, once the Fermi energy reaches a Landau level, it is locked on it for a large value of $N$ \cite{PRB.1993.47.9501}. When a degenerate Landau level in a subband is completely filled, the Fermi energy moves in the edge states as long as another Landau level is not reached.
By contrast, we already know that rotation eliminates the Landau-levels degeneracy.
As expected, there are no plateaus in the Fermi energy, instead, the behavior is approximately linear.
Moreover, Figs. \ref{figureenergy}(b)-(d) leads us to infer that the Fermi energy does not remain locked in a Landau level when a subband starts to
fill up. Instead, the Fermi energy oscillates between Landau states and edge states. In Fig \ref{FIG:fermi.N}(b),
we can observe an oscillatory behavior with both amplitude and period variables. The oscillations corresponds to the depopulation of a subband. For $\Omega=0$, the ranges where the Fermi energy has a linear behavior are almost self-explanatory. In effect, the Fermi energy is locked on a Landau level, which is evident when we observe the black dashed line. On the other hand, when $\Omega>0$, from the above discussion, the Fermi energy oscillates between Landau states and bulk states. As shown the Fig. \ref{FIG:fermi.N}(b), this causes changes in the linear behavior observed when $\Omega=0$.
It is evident that the edge states eliminate the abrupt change in the Fermi energy indicated by the black dashed line.
As can be seen from Figs. \ref{FIG:fermi.N}(a) and \ref{FIG:fermi.N}(b), the Fermi energy decreases with increasing angular velocity.
To explain these observations, we recall that one of the physical implications due to the rotation on the electronic states is to decrease the energy of states with angular number $m>0$. Figure \ref{FIG:fermi.N}(b) also shows that the rotation shifts all the oscillation peaks to the left.

\begin{figure}[!b]
\centering
\includegraphics[scale=0.23]{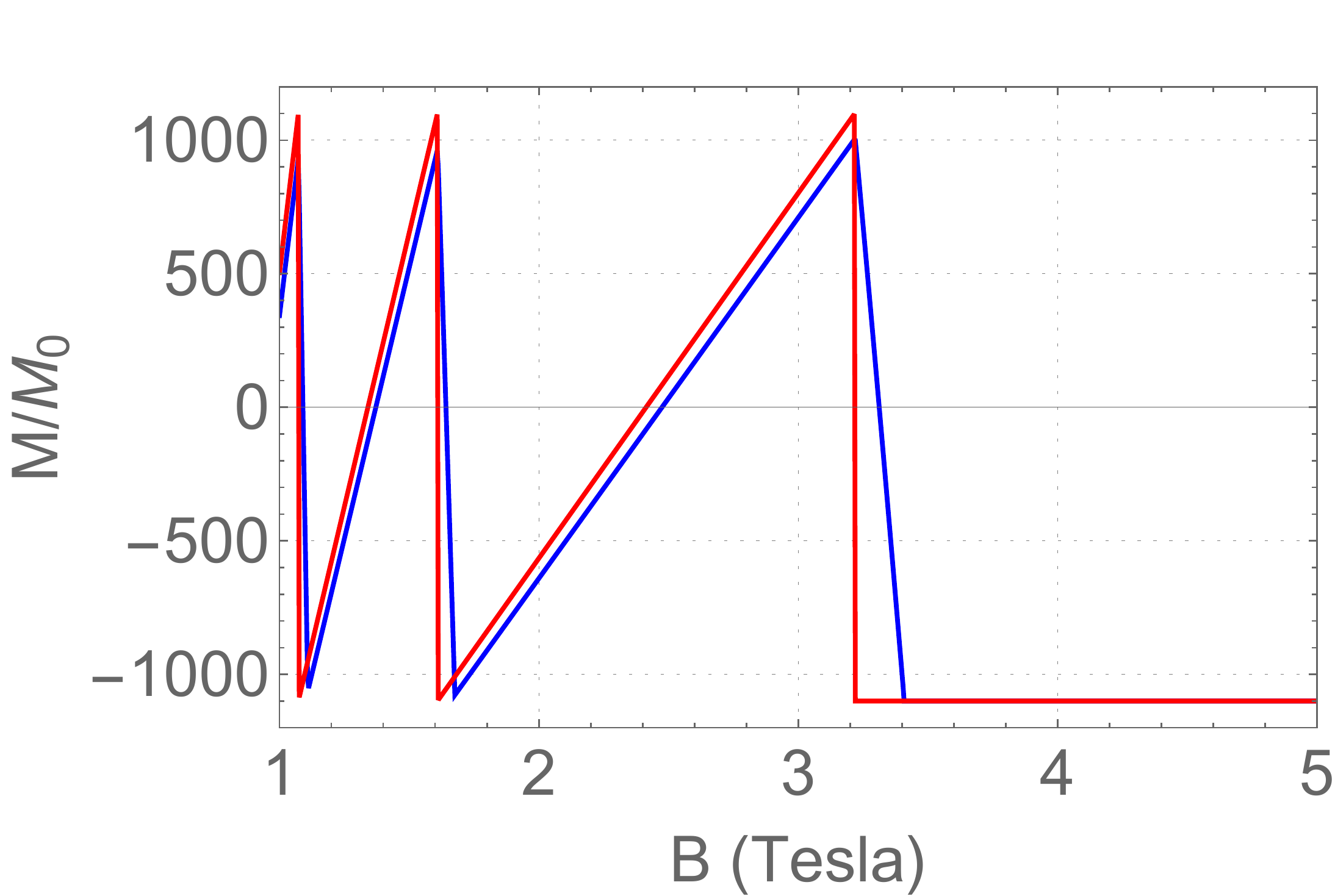}
\caption{Magnetization of a 2D ring (Eq. (\ref{Eq:Magnetizacao})) as function of the magnetic field .}
\label{FIG:Mag.1100E.Bulk}
\end{figure}

\subsection{Magnetization}

Now, we analyse the profile of the magnetization as a function of the magnetic field in a 2D quantum ring. At zero temperature, the magnetization is  derived from the relation
\begin{equation}
\mathcal{M}=-\frac{dU}{dB}, \label{Eq:Magnetizacao}
\end{equation}
where $U$ is the internal energy, which is expressed in terms of energy eigenvalues as
\begin{equation}
U=\sum_{\nu,m} E_{\nu,m}.
\end{equation}

The magnetic moment carries for each electronic state is given by
\begin{equation}
\mathcal{M}_{\nu,m}=-\frac{dE_{\nu,m}}{dB},
\label{Eq:momentomagnetico}
\end{equation}
and consequently the magnetization may be rewritten as
\begin{equation}
\mathcal{M}=\sum_{\nu,m} \mathcal{M}_{\nu,m}.
\end{equation}

Since the spectrum of the system is obtained numerically, we can not make use of Eq. (\ref{Eq:momentomagnetico}) to obtain the magnetic moment carried by a state. A way to overcome this limitation is to consider some approximations. We assume that the angular velocity is zero. The number of states in a subband can be computed from
\begin{equation}
N_{N}=\frac{\Phi}{\Phi_{0}},
\label{Eq:Nn}
\end{equation}
where $\Phi=\pi \left(r_{b}^{2}-r_{a}^{2}\right)B$ and $\Phi_{0}=h/e$ are the magnetic flux and magnetic flux quantum, respectively. We can estimate the number of edge states and energy of an edge state, respectively, as
\begin{equation}
N_{E}=n_{B}\frac{r_{b}}{\lambda},
\label{Eq:Ne}
\end{equation}
and
\begin{equation}
E_{\nu,m}=\left(\nu+\frac{1}{2}\right)\hbar \omega_{c}+i\Delta E,
\label{Eq:Enm.Edge}
\end{equation}
where $\Delta E \approx \hbar \omega_{c}/N_{E}$, $\nu=n=0,1,2,\dots$ and $1\leq i\leq N_{E}$. The Prefactor $n_{B}$
in Eq. (\ref{Eq:Ne}) counts the number of the edges. For a ring, which has two edges, $n_{B}=2$. On the other hand, for a quantum dot, $n_{B}=1$ \cite{PRL.1988.61.1001,Book.Yoshioka.2002}. The corresponding energy of a bulk state is obtained by making $i=0$ in Eq. (\ref{Eq:Enm.Edge}).

Using the results expressed by Eqs. (\ref{Eq:Nn})
and (\ref{Eq:Enm.Edge}), we can show that the internal energy can be written as
\begin{align}
U&= \hbar \omega_{c}\Bigg\{\left[ \frac{p^{2}}{2}+\left( p+\frac{1}{2}\right) q\right] N_{N}+ \frac{p}{2}\left( N_{E}+1\right) \notag \\
&+\frac{\left[ N_{e}+N_{n}\left(
q-1\right)\right]\left[N_{e}+N_{n}\left(
q-1\right)+1\right]}{2N_{E}}\Bigg\} ,
\label{Eq:EnergiaInterna}
\end{align}
where $p=N/N_{N}$ and $q=N/N_{N}-p$. Consequently, using the Eq. (\ref{Eq:Magnetizacao}), the magnetization reads
\begin{align}
\mathcal{M}=&-\mathcal{M}_{0}\Bigg\{\left(2p+1\right)N- 2p\left(p+1\right)N_{N}\notag \\
&
+ p\left( \frac{3}{2}N_{E}+1\right) +\Bigg[1+\left(q-1\right)\frac{N_{N}}{N_{E}}\Bigg]\notag \\
&\times
\Bigg[\frac{3}{2}\left(N_{E}+1\right)+ \frac{5}{2}\left(q-1\right)N_{N}-2N\Bigg]\notag \\
& - \frac{N}{N_{E}}-\frac{1}{2}
\Bigg\},
\label{Eq:Magnetizacao}
\end{align}
where $\mathcal{M}_{0}=\hbar e/2\mu$.
\begin{figure}[!h]
\centering
\includegraphics[scale=0.32]{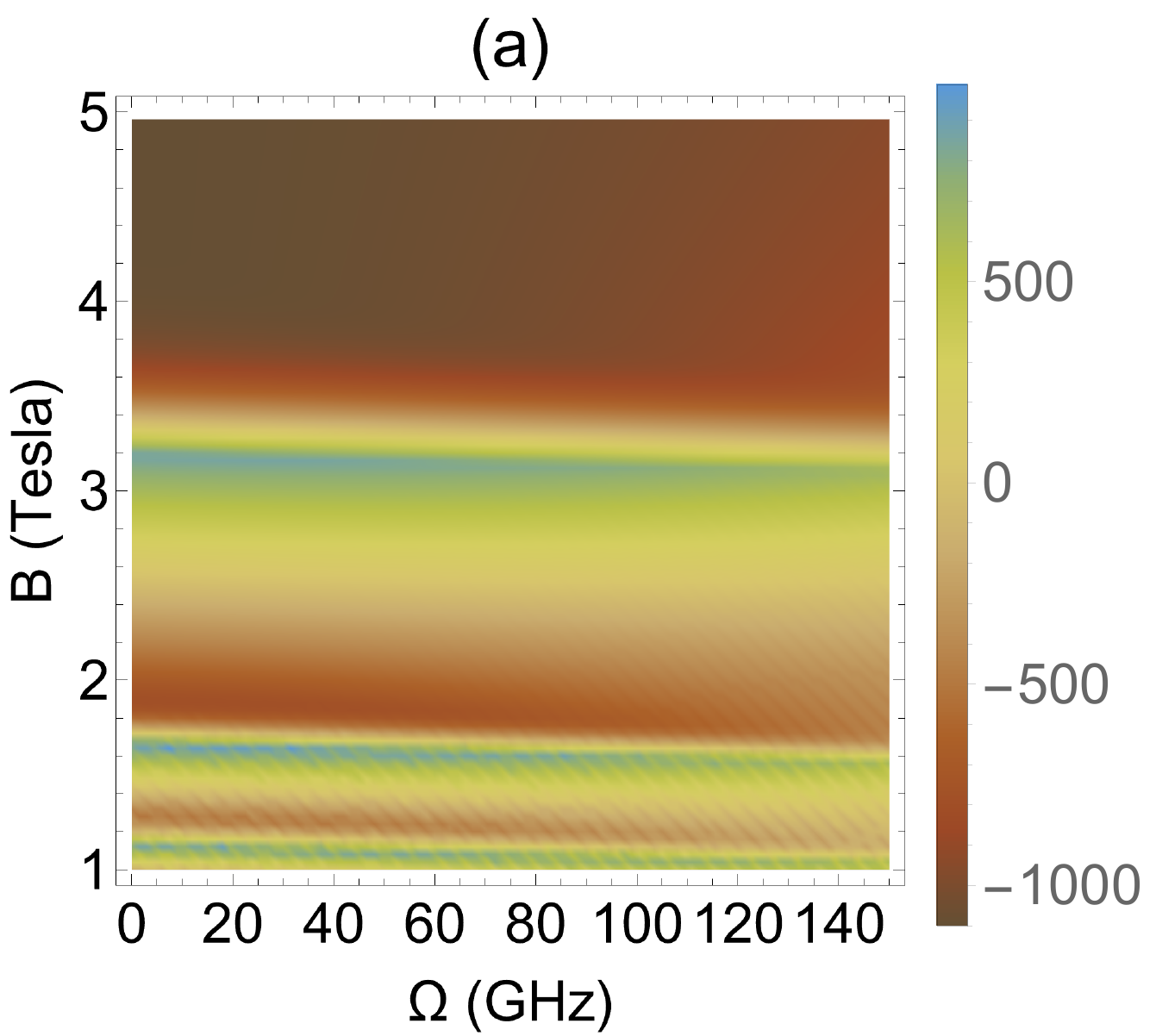}
\includegraphics[scale=0.32]{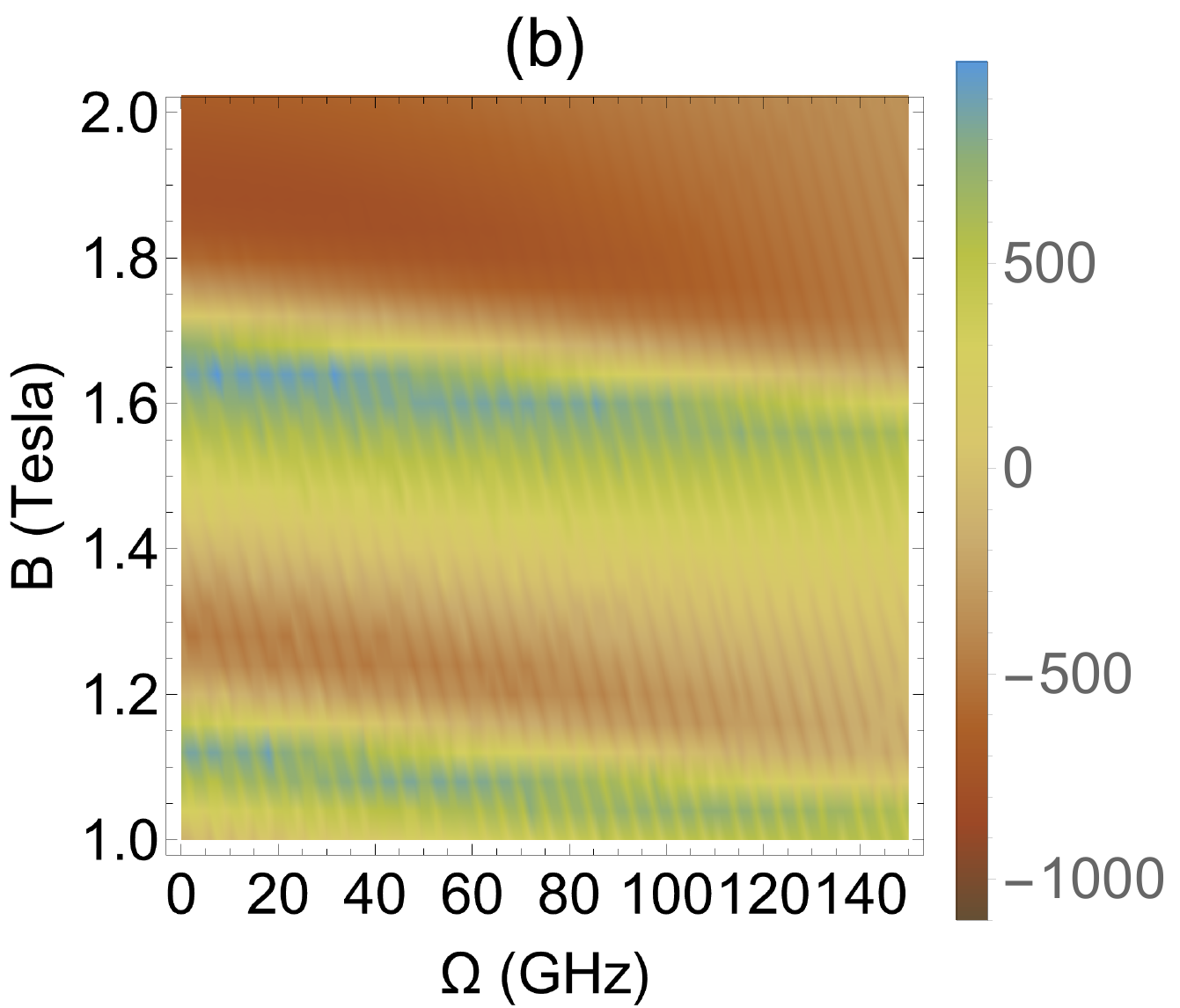}
\caption{Magnetization of a 2D ring (Eq. (\ref{Eq:Mag.aproximado})) as a function of magnetic field and angular velocity.}
\label{FIG:Mag.1100E.DP}
\end{figure}

The magnetization in a $2$D quantum ring as a function of magnetic field given by Eq. (\ref{Eq:Magnetizacao}) is shows in Fig. \ref{FIG:Mag.1100E.Bulk}. The red solid line describes the usual bulk oscillations of the magnetization in a 2D quantum ring. These oscillations are called dHvA oscillations and result from the depopulation of a subband in the Fermi energy. Each abrupt change in the magnetization corresponds to a jump of the Fermi energy from one level to the next lower level. The solid blue line shows the effect of the edge states in the dHvA oscillations. As we can observe, the edge states remove the discontinuous jumps in the magnetization. In fact, now the density of states is not composed of $\delta$ functions. Besides, the amplitude of the oscillations decreases. When the magnetic field is strong enough to make only one occupied subband, the magnetization tends to the value $-N\mathcal{M}_{0}$.

Equation (\ref{Eq:Magnetizacao}) shows as the edge states affect the behavior of the dHvA oscillations without taking into account rotating effects.
As mentioned above, the energies are determined numerically by using the transcendental equation (\ref{eq.transcendental}), and it is not easy to get accurate physics information from the model. An alternative method to overcome this is to calculate the magnetization with small steps $\Delta B$. The approximate expression is
\begin{equation}
\mathcal{M} \approx-\frac{\Delta U}{\Delta B}.
\label{Eq:Mag.aproximado}
\end{equation}

Figure \ref{FIG:Mag.1100E.DP} illustrates a density plot as a function of the magnetic field and angular velocity of a 2D quantum ring by using the approximate formula  (\ref{Eq:Mag.aproximado}). In Fig. \ref{FIG:Mag.1100E.DP}(a), we can observe an oscillatory pattern along a line with constant angular velocity. These oscillations correspond to the dHvA effect already observed when we discuss the magnetization in Fig. \ref{FIG:Mag.1100E.Bulk}. As we can see, the amplitude of these oscillations increases with an increasing of the magnetic field. Such an effect is associated with the presence of edge states.

Rotating effects are also evident in Fig. \ref{FIG:Mag.1100E.DP}(a). As we can clearly see, the rotation shifts the dHvA oscillations to lower magnetic fields. In addition to this, rotation decreases the amplitude of the dHvA oscillations. In the absence of rotating effects ($\Omega=0$), we can observe that the position of the valleys and peaks of the dHvA oscillations in Fig. \ref{FIG:Mag.1100E.DP}(a) are in very good agreement with those observed in Fig. \ref{FIG:Mag.1100E.Bulk}.
For strong magnetic fields, the magnetization tends much
faster to the value $-N\mathcal{M}_{0}$ when $\Omega \rightarrow 0$.

Besides the dHvA oscillations, AB-type oscillations also are observed along a line with constant angular velocity. These oscillations are more evident when we consider a smaller magnetic field range, as shown in Fig. \ref{FIG:Mag.1100E.DP}(b). The origin of these oscillations is related to the crossing of the edge states at the Fermi energy, as we can infer from Fig. \ref{Fig:EnmxB}. Figure \ref{FIG:Mag.1100E.DP}(b) also shows that that AB-type oscillations are also present along a line with constant magnetic field. In other words, the rotation induces AB-type oscillations, which is an expected result. Indeed, as can see in Fig. \ref{Fig:EnmxW}, the rotation induces energy level crossings.
We emphasize that the nearly periodic behavior of AB-type oscillations is related to a small number of occupied subbands.

\section{Conclusions}
\label{Sec:conclusoes}

In summary, we have studied the physical implications due to rotation effects on the electronic structure of a two-dimensional quantum ring immersed in a uniform magnetic field considering edge effects. Initially, we addressed the problem of an electron in a region without confinement potential. This initial approach illustrated many aspects of the effects of rotation on Landau levels. The energy eigenvalues are written in terms of a minimum potential energy. The minimum potential provides important information about the energy that the electron acquires in a given state of the system, since it depends on the quantum number $m$, and the cyclotron and rotation frequencies. Among the various results investigated,
we point out that the rotation lifts the Landau-levels degeneracy.

Subsequently, the confinement of electrons in a quantum ring was considered. To describe the quantum ring, we consider the hard wall potential model. Due to the shape of the confinement potential, states inside the ring and far away from the edges do not feel the potential. The situation is then similar to that studied in Section \ref{sec2}. On the other hand, close to edges, electronic states are significantly influenced by the walls. These states are known as edge states. We verified that they play an important role in Fermi energy and magnetization. The importance of edge states in physical phenomena has been reported in other works, for example in Refs. \cite{PRB.1982.25.2185,PRL.1988.61.1001,PRB.1993.47.9501}.

The effects of rotation on Fermi energy and magnetization are also evident in our analyses.
One of the most important results in our investigation has revealed that
the rotation induces AB-type oscillations at magnetization. At this point, we must remember that Vignale and Mashhoon \cite{PLA.1995.197.444} had already verified that rotation induces another equilibrium property, specifically, persistent current. The results we have reported here emphasize the importance of investigations of physical phenomena in mesocopic systems set in rotation.

In the last decades, there has been a growing study on the feasibility of rotating nanostructures \cite{RP.2011.1.17}.
For instance,
Král and Sadeghpour showed that circularly polarized light generates a rotary motion of carbon nanotubes \cite{PRB.2002.65.161401}. They also discuss possible applications to rotating microdevices. Another example is given in
Ref. \cite{Nature.2003.424.408}, where the authors  report on the construction and successful operation of a fully synthetic nanoscale electromechanical actuator incorporating a rotatable metal plate, with a multi-walled carbon nanotube serving as the key motion-enabling element. We must also emphasize the rapid progress in manufacturing, controlling and understanding structures in the mesoscopic regime \cite{BOOK.Heinzel.2008}. Thus, we have a favourable environment for the experimental investigation of the physical properties of mesoscopic systems in a rotating frame can be performed.
Finally, we have considered high rotational frequencies of the order of $10^{9}$ in order to observe rotation effects. This detail was mentioned in Refs. \cite{PLA.1995.197.444,PRB.1996.54.1877} in which the authors show that much higher rotational frequencies are required to produce appreciable quantum rotational effects.

\section*{Acknowledgments}

This work was partially supported by the Brazilian agencies CAPES, CNPq and
FAPEMA. E. O. Silva acknowledges CNPq Grants 427214/2016-5 and 303774/2016-9, and FAPEMA Grants 01852/14.

\bibliographystyle{model1a-num-names}

\end{document}